\documentclass{elsarticle}

\usepackage{amsmath}
\usepackage{amssymb}
\usepackage{bm}
\usepackage{cases}
\usepackage{color}
\usepackage{float}
\usepackage{geometry}
\usepackage{graphics}
\usepackage{hyperref}
\usepackage{multicol}
\usepackage{multirow}
\usepackage{setspace}

\usepackage{subfig}

\biboptions{sort&compress}

\graphicspath{{FIG/}}

\begin{document}
\title{A discrete unified gas-kinetic scheme for immiscible two-phase flows}
\author{Chunhua Zhang}
\author{Kang Yang}
\author{Zhaoli Guo\corref{COR}}
\ead{zlguo@hust.edu.cn}
\cortext[COR]{}
\address{State Key Laboratory of Coal Combustion, School of Energy and Power Engineering, Huazhong University of Science and Technology, Wuhan 430074, China}

\begin{abstract}
In this work, we extend the discrete unified gas-kinetic scheme (DUGKS) [Guo \emph{et al}., Phys. Rev. E \textbf{88}, 033305 (2013)] to continue two-phase flows.
In the framework of DUGKS, two kinetic model equations are used to solve
the quasi-incompressible phase-field governing equations [Yang \emph{et al}., Phys. Rev.E \textbf{93}, 043303 (2016)].
One is for the Chan-Hilliard (CH) equation and the other is for the Navier-Stokes equations. The DUGKS can correctly recover the quasi-incompressible phase-field governing equations through the Chapman-Enskog analysis.
Unlike previous phase-field-based LB models, the Courant-Friedricks-Lewy condition in DUGKS is ajustable which can increase numerical stability. Furthermore, with the finite-volume formulation the model can be easily implemented on non-uniform meshes which can improve numerical precision.
The proposed model is validated by simulating a stationary drop, layered Poiseuille flow, rising bubble and Rayleigh-Taylor instability and comparing with the quasi-incompressible lattice Boltzmann method (LBM).
Numerical results show that the method can track the interface with high accuracy and stability.
The  model is also capable of dealing with a wider range of viscosity and density ratios than the quasi-incompressible lattice Boltzmann model.  The present model is a promising tool for numerical simulation of two-phase flows.

keywords: Multiphase flow; Finite-volume method; Discrete unified gas-kinetic scheme; Lattice Boltzmann method; Non-uniform gird
\end{abstract}
\maketitle

\section{Introduction}

Recently, modeling multiphase flows based on kinetic descriptions has received particular attention. In kinetic schemes, intermolecular interactions that determine phase behaviors are incorporated
at the mesoscopic level into a discretized Boltzmann equation such that the complex macroscopic fluid behaviour, such as phase separation or coalescence, is a result of intermolecular interactions.
This feature brings some distinct advantages, such as free of interface tracking.
The most popular kinetic method for two-phase flows may be the lattice Boltzmann Equation (LBE) method, which solves the discrete velocity Boltzmann equation (DVBE) on a regular lattice
\cite{rk1991,rk1993,swift1995,swifttwo1995,heshan1998,hechen1999}.
To date, the LBE method has achieved much success in the study of multiphase flows \cite{heshan1998,sc1994,wangy2015,liuhh2016}.

However, most existing  multiphase LBE models share some undesired features in practical applications, such as the numerical instability in simulating systems with high density and viscosity ratios, although some efforts have been made from different viewpoints \cite{lee2005,inamuro2004, inamuro2016,zheng2005,wangy2015}.
Another inconvenience is that most LBE models are designed based on isotropy and uniform grid in order to perfect shift in a single time step.
This treatment simplifies the algorithm greatly but may cause some difficults for certain problems requiring non-uniform meshes.

Recently, a new type of kinetic method, discrete unified gas kinetic scheme (DUGKS)  is proposed for single phase multiscale flows \cite{guo2013}. It has been shown that, even for continuum flows, the DUGKS also has some distinct features that distinguish it from other kinetic schemes. Generally, the features of the DUGKS can be summarized as follows. First,
DUGKS is a finite volume scheme which is easy to perform on irregular meshes \cite{zhu2016}. Second, the DUGKS couples the streaming and collision processes for flux evaluation, which guarantees a low numerical dissipation~\cite{wang2015,guo2013}.
Thirdly, the mesh size and time step in DUGKS are decoupled, such that the time step is determined independently by the Courant-Friedrichs-Lewy (CFL) condition.
These features make it possible to extend the DUGKS to two-phase flows with improved properties in comparison with LBE model, and this is the aim of the present study.

The rest of this paper is organized as follows. In Sec.~II, the methodology of the proposed DUGKS for two-phase flows will be introduced, and in Sec.~III, several numerical tests are carried out to demonstrate the performance of the proposed method. Finally, a brief summary is presented in Sec.~IV.

\section{METHODOLOGY}
\subsection{Quasi-incompressible phase-field model }
In the phase-field theory for a binary fluid system, the thermodynamic behavior is described by a free-energy function related to an order parameter $\phi$ and its spatial derivatives. The order parameter is used to distinguish the different fluids and varies continuously from one fluid to the other fluid.  
A Landau free-energy function is defined as
\begin{equation}\label{freeenergyfunction}
  F(\phi)=\int_\Omega\left[\psi(\phi)+\frac{\kappa}{2}|\nabla \phi|^2 \right]d\Omega,
\end{equation}
where $\psi(\phi) $ is the bulk free-energy density, $\kappa$ is the coefficient of the surface tension, and $\Omega$ is the control volume.
For an isothermal system, the following double-well form of free-energy density \cite{jacqmin1999,jacqmin2000} can be used,
\begin{equation}\label{doublewell}
  \psi(\phi)=\beta(\phi-\phi_A)^2(\phi-\phi_B)^2,
\end{equation}
where $\phi_A$ and $\phi_B$ are constants corresponding to the equilibrium state of the order parameters, i.e., the order parameters to mark the bulk fluids A and B, respectively. $\beta$ is a constant that controls the interfacial thickness $W$ \cite{jacqmin1999,jacqmin2000,yan2007} and the surface tension force $\sigma$ \cite{yan2007},
\begin{align}
\label{thickness}
  W & =\frac{1}{\phi_A-\phi_B}\sqrt{\frac{8\kappa}{\beta}},\\
\label{sigma}
  \sigma & =\frac{|\phi_A-\phi_B|^3}{6}\sqrt{2\kappa\beta}.
\end{align}
The variation of the free-energy function $F(\phi)$ with respect to the order parameter yields the chemical potential \cite{jacqmin1999,jacqmin2000,yan2007},
\begin{equation}\label{chemicalpotential}
  \mu_\phi =\frac{\delta F}{\delta \phi}=\frac{\partial \psi}{\partial \phi}- \kappa\nabla^2\phi   \\
  =4\beta(\phi-\phi_A)(\phi- \phi_B)(\phi-\frac{\phi_A+\phi_B}{2})-\kappa\nabla^2\phi.
\end{equation}
For a flat interface in an equilibrium state, the equilibrium interface profile can be obtained by solving $\mu_{\phi}(\phi)=0$. The order-parameter profile across the interface is represented as
\begin{equation}\label{equilibriuminterface}
\phi(\zeta)=\frac{\phi_A-\phi_B}{2}+\frac{\phi_A-\phi_B}{2}\tanh\left(\frac{2\zeta}{W}\right),
\end{equation}
where $\zeta$ is the signed distance in the direction normal to the interface.
The interface evolution can be described in terms of the order parameter governed by the convective Cahn-Hilliard (CH) equation \cite{cahn1958,cahn1959,jacqmin1999},
\begin{equation}\label{CH0}
  \frac{\partial \phi}{\partial t}+\nabla\cdot(\phi\bm u)=\nabla\cdot(\lambda\nabla\mu_{\phi}),
\end{equation}
where $\bm u$ is the velocity and $\lambda$ is the mobility.

For most existing multiphase LBM models, the fluid is usually assumed to be incompressible in the whole domain, i.e., $\nabla\cdot \bm{u}=0$. However, this assumption leads to the violation of mass conservation as the two fluids have different densities \cite{chella1996,lowengrub1998,guo2014,shen2013}. To overcome this problem,  a quasi-incompressible model that conserves mass locally was  developed \cite{yang2016}, which will be adopted in the present study. In this model, the hydrodynamic equations for a two-phase system are given by,
\begin{align}\label{Quasimass}
  \nabla\cdot\bm u &=-\gamma\nabla\cdot \lambda\nabla\mu_{\phi},\\
\rho \left(\frac {\partial \bm u}{\partial t}+ \bm u\cdot\nabla \bm u \right) &=-\nabla p+ \nabla\cdot \left[\rho\nu(\nabla \bm{u}+ \nabla\bm{u}^T) \right]+\bm F,
\end{align}
with
\begin{equation}\label{oderrho}
  \rho=\frac {\phi-\phi_B}{\phi_A-\phi_B}\rho_A  + \frac{\phi_A-\phi}{\phi_A-\phi_B}\rho_B,
\end{equation}
\begin{equation}\label{Gama}
  \gamma=\frac{\rho_r-1}{\phi_A-\phi_B\rho_r},
\end{equation}
where $\rho_A$ and $\rho_B$ are the densities of the two fluids, respectively, $\bm F$ is the total force, including the interfacial force $\bm F_s (=-\phi\nabla\mu_{\phi})$ and other body forces $\bm F_b$, such as gravity, $p$ is the hydrodynamic pressure, $\nu$ is the kinematic viscosity, and $\rho_r=\rho_A/\rho_B$ is the density ratio.

It is clear that the velocity field is no longer divergence-free and the fluid is compressible in the mixing zone. Substituting Eq. (\ref{Quasimass}) into (\ref{CH0}), one can obtain,
\begin{equation}\label{locallyconserved}
  \partial_t \rho +\nabla\cdot(\rho \bm u)=0,
\end{equation}
which means that the mass is conserved locally in this model.

\subsection{ DUGKS for two-phase flows}
Based on the Boltzmann-BGK equation, Guo~\emph{et al.} \cite{guo2013} developed a type of discrete unified gas kinetic scheme (DUGKS) by combining the advantages of both LBE and unified gas kinetic scheme methods (UGKS) \cite{xu2010}. The starting point of the original DUGKS is the Boltzmann equation with BGK collision model \cite{bgk1954}.
Now we extend the scheme to two-phase flows described by the quasi-incompressible phase-field model described above. 
To this end, we adopt the following kinetic model \cite{yang2016} as the starting point,
\begin{equation}\label{bgkf}
\frac{\partial f_i}{\partial t}+ \bm{\xi}_i \cdot\nabla f_i =
-\frac{f_i-f_i^{eq}}{\tau_f}+F^{f}_i,
\end{equation}
\begin{equation}\label{bgkg}
\frac{\partial g_i}{\partial t}+ \bm{\xi}_i \cdot\nabla g_i =
-\frac{g_i-g_i^{eq}}{\tau_g}+F^{g}_i,
\end{equation}
where $f_i\equiv f_i(\bm{x},\bm{\xi}_i,t)$ and $g_i\equiv g_i(\bm{x},\bm{\xi}_i,t)$ are the particle distribution function (DF) with  discrete velocity $\bm{\xi}_i$ at position $\bm x$ and  time $t$ for the hydrodynamics and order parameter fields, respectively, the subscript $i$ is the lattice velocity direction, $f_i^{eq}$ and $g_i^{eq}$ are the corresponding equilibrium distribution functions (EDF),
 $\tau_{f}$ and $\tau_g$ are the corresponding relaxation time for different distribution functions, $F_i^{f}$ and $ F_i^{g}$ are the source terms.  Here $f_i$ is a new variable introduced to change the particle distribution function for density and momentum into that for pressure and momentum. Detailed information about the transformation process can be found in \cite{hechen1999,lee2005}.
 The macroscopic variables are given by the first two moments of the DFs,
 \begin{equation}\label{phiup}
\phi(\bm x,t)=\sum_{i=0}^{Q-1} g_i(\bm{x},\bm{\xi}_i,t), \quad p(\bm x,t)=\sum_{i=0}^{Q-1} f_i(\bm{x},\bm{\xi}_i,t),\quad  RT\rho(\bm x,t)\bm u=\sum_{i=0}^{Q-1} \bm{\xi}_i f(\bm x,\bm{\xi}_i,t),
 \end{equation}
 where $Q$ denotes the number of discrete velocities and $Q$ is set to be $9$ in this study, $T$ is a constant temperature, and $R$ is the gas constant.
 The density can be obtained by the Eq.~(\ref{oderrho}).
 The choice of EDF  $f_i^{eq}$ must satisfy the conservation of momentum while the choice of EDF  $g_i^{eq}$ must satisfy the conservation of order parameter.
 It can be shown that the kinetic model given by  Eqs.~(\ref{bgkf}) and (\ref{bgkg}) can recover the quasi-incompressible phase-field model described by Eqs.~(\ref{CH0}-\ref{Gama}).

For convenience, we rewrite Eq. (\ref{bgkf}) and Eq. (\ref{bgkg}) in the following form,
\begin{equation}\label{bgkvar}
\frac{\partial \varphi_i}{\partial t}+ \bm{\xi}_i \cdot\nabla \varphi_i = \Omega_i^{\varphi}+ F^{\varphi}_i,
\end{equation}
where $\varphi=f \text{or} g$, and $\Omega_i^{\varphi}\equiv-(\varphi_i-{\varphi}_i^{eq})/\tau_{\varphi}$.
In DUGKS, the flow domain is divided into a set of control volumes (cells). Integrating Eq. (\ref{bgkvar}) over a control volume $V_j$ centered at $\bm x_j$ from $t_{n}$ to $t_{n+1}$
(the time step $\Delta t=t_{n+1}-t_n$ is assumed to be a constant in the present work), and using the midpoint rule for the
integration of the flux term at the cell interface and trapezoidal rule for the collision and source terms inside each cell, one can obtain
\begin{equation}\label{integrationOne}
\varphi_i^{n+1}-\varphi_i^{n}+\frac{\Delta t}{|V_j|}J^{n+1/2}=\frac{\Delta t}{2} [\Omega_i^{\varphi,n+1}+\Omega_i^{\varphi,n}]+\frac{\Delta t}{2}[F_i^{\varphi,n+1}+F_i^{\varphi,n}],
\end{equation}
where
\begin{equation}\label{microfluxf}
 J^{n+1/2}=\int_{\partial V_j}\left(\bm{\xi}_i \cdot \bm n\right)\varphi_i\left(\bm x_j,\bm{\xi}_i,t_{n+1/2}\right)d{\bm S} ,
\end{equation}
is the flux across the cell interface,  $|V_j|$ and $\partial V_j$ are the volume and surface area of cell $V_j$, $\bm{n}$ is the outward unit vector normal to the surface.
It is noted that $\varphi_i$ and $F_i^{\varphi}$ are changed to the cell-averaged values of the distribution function and source term located in the control volume $V_j$, i.e.,
\begin{equation}\label{aveage}
  \varphi_i^n=\frac{1}{|V_j|} \int_{V_j}{\varphi_i(\bm{x}_j,\bm{\xi}_i,t_n)}d\bm{x}, \quad F_i^{\varphi,n}=\frac{1}{|V_j|}\int_{V_j}{ F_i^{\varphi}(\bm{x}_j,\bm{\xi}_i,t_n)}d\bm{x}.
\end{equation}
 It is clear that Eq.~(\ref{integrationOne}) is implicit since $\varphi_i^{eq}$ in the collision term $\Omega_i^{\varphi,n+1}$ involves the unknow macroscopic conserved variables at $t_{n+1}$. In order to remove the implicity, two auxiliary distribution functions are introduced,
\begin{subequations}\label{auxf}
\begin{align}
\tilde{\varphi}_i & =\frac{2\tau_{\varphi}+\Delta t}{2\tau_{\varphi}}\varphi_i-\frac{\Delta t}{2\tau_{\varphi}}\varphi_i^{eq}-\frac{\Delta t}{2}F^{\varphi}_i, \\
\tilde{\varphi}^+_i & =\frac{2\tau_{\varphi}-\Delta t}{2\tau_{\varphi}+\Delta t}\tilde{\varphi}_i+\frac{2\Delta t}{2\tau_{\varphi}+\Delta t}\varphi_i^{eq}+\frac{2\tau_{\varphi} \Delta t}{2\tau_{\varphi}+\Delta t}F^{\varphi}_i.
\end{align}
\end{subequations}
Substituting Eq. (\ref{auxf}) into Eq. (\ref{integrationOne}), we can obtain
\begin{equation}\label{updatef}
\tilde{\varphi}^{n+1}_{i}=\tilde{\varphi}^{+,n}_{i}-\frac{\Delta t}{|V_j|}J_{\varphi}^{n+1/2}.
\end{equation}

Based on the Eq.~(\ref{auxf}a), the conserved variables can be computed from $\tilde {\varphi}$,
\begin{equation}\label{conservedVariable}
\begin{aligned}
\phi=\sum_{i}^{Q} \tilde {g_i}+\frac{\Delta t}{2}\sum_{i}^{Q} F_i^{g},\quad p=\sum_{i}^{Q} \tilde {f_i}+\frac{\Delta t}{2}\sum_{i}^{Q} F_i^{f},\\
RT\rho\bm u=\sum_{i}^{Q} \bm{\xi}_i \tilde{f_i}+\frac{\Delta t}{2}\sum_{i}^{Q} \bm{\xi}_i F_i^{f}.
\end{aligned}
\end{equation}
Therefore, in practical simulations, we only need to track the distribution function $\tilde{\varphi}$ instead of the original one.

The key ingredient in updating $\tilde{\varphi}_i^{n+1}$ is to evaluate the interface flux $\bm{J}^{n+1/2}$. According to Eq~(\ref{microfluxf}), it is clear that the interface flux is only determined by the original distribution function $\varphi_i(\bm{x}_j,\bm{\xi}_i,t_{n+1/2})$ at the half time step. Similar to the treatment in Eq. (\ref{integrationOne}), we integrate the Eq. (\ref{bgkvar})  within a half time step $h=\Delta t/2$ along the characteristic line with the end point located at the cell interface $(\bm{x_b}=\bm{x}_j+\bm{\xi}_i h)$,
 \begin{equation}\label{integrationHalf}
\begin{aligned}
\varphi_i(\bm{x_b},\bm{\xi}_i,t_n+h)-\varphi_i(\bm{x_b}-\bm{\xi}_i h,\bm{\xi}_i,t_n)  = \frac{h}{2} \left[\Omega_i^{\varphi}(\bm{x_b},\bm{\xi}_i,t_n+h)+\Omega_i^{\varphi}(\bm{x_b}-\bm{\xi}_i h,\bm{\xi}_i,t_n)\right] \\
+ \frac{h}{2}\left[F_i^{\varphi}(\bm{x_b},\bm{\xi}_i,t_n+h)+F_i^{\varphi}(\bm{x_b}-\bm{\xi}_i h,\bm{\xi}_i,t_n)\right].
\end{aligned}
\end{equation}

To remove the implicity, another two auxiliary  distribution functions $\bar{\varphi}_i$ and $\bar{\varphi}^+_i$ are introduced
\begin{subequations}\label{auxfbar}
\begin{align}
\bar{\varphi}_i &=\frac{2\tau_{\varphi}+ h}{2\tau_{\varphi}}\varphi_i-\frac{h}{2\tau_{\varphi}} \varphi_i^{eq}-\frac{h}{2}F^{\varphi}_i, \\
\bar{\varphi}^+_i &=\frac{2\tau_{\varphi}- h}{2\tau_{\varphi}+ h}\bar{\varphi}_i  +\frac{2 h}{2\tau_{\varphi}+ h}\varphi_i^{eq}+\frac{2\tau_{\varphi}  h}{2\tau_{\varphi}+ h}F^{\varphi}_i.
\end{align}
\end{subequations}
As a result, Eq.(\ref{integrationHalf}) can be rewritten in an explicit formulation,
\begin{equation}\label{fbar}
\bar{\varphi}_i(\bm{x}_b,\bm{\xi}_i, t_n+h)=\bar{\varphi}_i^+(\bm{x}_b-\bm{\xi}_i h,\bm{\xi}_i,t_n).
\end{equation}

With the Taylor expansion around the cell interface $\bm x_b$, for smooth flows, $\bar{\varphi}_i^+(\bm{x}_b-\bm{\xi}_i h,\bm{\xi}_i,t_n)$ can be approximated as
\begin{equation}\label{fbar_h}
 \bar{\varphi}_i^+(\bm{x}_b-\bm{\xi}_i h,\bm{\xi}_i,t_n)=\bar{\varphi}_i^+(\bm{x}_b,\bm{\xi}_i,t_n)-\bm{\xi}_i h\cdot \bm{\sigma}_b ,
\end{equation}
 where $\bm{\sigma}_b=\nabla \bar{\varphi_i}^+(\bm x_b,\bm{\xi}_i,t_n)$ and the gradient term can be approximated by linear interpolations.
 Once the distribution function $\bar{\varphi}_i$ is updated, the macroscopic variables ($\phi, \bm u, p$) at the cell interface can be obtained by replacing  $h$ with $\Delta t$ and  $\bar{\varphi}_i$ with $\tilde{\varphi}_i $ in Eq.~(\ref{conservedVariable}). Thus, the equilibrium distribution function $\varphi^{eq}(\bm{x}_b,\bm{\xi}_i,t_n+h)$ can be calculated by the macroscopic variables at the cell interface which will be shown later.
From Eq.~(\ref{auxfbar}a), the original distribution function becomes
\begin{equation}\label{original-function}
  \varphi_i=\frac{2\tau_{\varphi}}{2\tau_{\varphi}+h}\bar{\varphi_i}+ \frac{h}{2\tau_{\varphi}+h}\varphi_i^{eq}+\frac{\tau_{\varphi} h}{2\tau_{\varphi}+h} F^{\varphi}_i.
\end{equation}
As a result, the micro-flux  $\bm{J}^{n+1/2} $ can be calculated through the Eq.~(\ref{microfluxf}).
Moreover, according to Eqs.~(\ref{auxf}) and (\ref{auxfbar}), the following relations are easily established by algebra calculation,
\begin{equation}\label{relations1}
\bar{\varphi}_i^+=\frac{2\tau_{\varphi}-h}{2\tau_{\varphi}+\Delta t}\tilde{\varphi}_i+ \frac{3 h}{2\tau_{\varphi}+\Delta t}\varphi^{eq}_i+\frac{3\tau_{\varphi}h}{2\tau_{\varphi}+\Delta t}F^{\varphi}_i,
\end{equation}
\begin{equation}\label{relations2}
\tilde{\varphi}_i^+=\frac{4}{3}\bar{\varphi}_i^+ -  \frac{1}{3}\tilde{\varphi}_i.
\end{equation}
 In the end, the distribution function $\tilde \varphi_i^{n+1}$ is updated according to Eq. (\ref{updatef}).
Note that the time step $\Delta t$ is an adjustable variable in the DUGKS and only determined by Courant-Friedrichs-Lewy (CFL) condition,
 \begin{equation}\label{cfl}
 \Delta t=\alpha \frac{\Delta x}{C_{max}},
 \end{equation}
 where $\alpha$ is the $\mbox{CFL}$ number and lies between $0$ and $1$, $C_{max}$ is in the order of the maximal discrete velocity and $\Delta x$ is the minimal grid spacing.

 In the present study, Both uniform and nonuniform meshes are considered. The two-dimensional and nine velocity  discrete model is employed in both DUGKS and LBE models, which is generated using the tensor product method~\cite{guo2013,wuchen2016}. And the discrete velocities $\bm {\xi}_i$ can be written as
 \begin{equation}\label{D2Q9}
   \bm{\xi}_i=
   \begin{cases}
   (0,0)c, \hspace{2mm}  & \mbox{i =0} \\
   (\cos[(i-1)\pi/2],\sin[(i-1)\pi/2])c, \hspace{2mm}  & \mbox{i  =1,\ldots,4}\\
  \{\cos[(i-5)\pi/2+\pi/4],\sin[(i-5)\pi/2+\pi/4]\}\sqrt{2}c, \hspace{2mm}  & \mbox{i =5,\ldots,8},
   \end{cases}
 \end{equation}
 where $c=\sqrt{3RT}$. In order to recover the quasi-incompressible phase-field governing equations, the equilibrium distribution functions $f_i^{eq}$ and $g_i^{eq}$  are respectively defined as
\begin{align}\label{feq}
f^{eq}_i & =\omega_i p+c_s^2\rho s_i(\bm{u}),\\
\label{geq}
g^{eq}_i & =H_i+ \phi s_i(\bm{u}),
\end{align}
with
\begin{equation}\label{su}
s_i (\bm u)=\omega_i \left[\frac{\bm{\xi}_i\cdot\bm u}{c_s^2}+  \frac{(\bm{\xi}_i \cdot\bm{u})^2}{2 c_s^4}- \frac{\bm{u}^2}{2 c_s^2}\right],
\end{equation}
\begin{equation}\label{Hi}
H_i=
\begin{cases}
\phi-(1-\omega_0)\eta\mu_{\phi},              & i=0 \\
\omega_i\eta\mu_{\phi},              & i\neq0
\end{cases}
\end{equation}
where $\eta$ is an adjustable parameter for a given mobility, $\omega_i$ is the weighting coefficients which are defined as $\omega_0=4/9,~\omega_{1,\ldots,4}=1/9$ and $\omega_{5,\ldots,8}=1/36$.
The source terms $F^f_i$ and $F^g_i$ are defined as
\begin{equation}\label{F}
F^{f}_i=(\bm{\xi}_i -\bm{u})\cdot [\bm{F}\varGamma _i (\bm u)+s_i (\bm{u}) c_s^2\nabla \rho]-\omega_i c_s^2\rho\gamma\nabla\cdot  (\lambda\nabla\mu_{\phi}),
\end{equation}
\begin{equation}\label{G}
 F^{g}_i= \frac{\phi}{c_s^2\rho}( \bm{\xi}_i - \bm{u})\cdot(\bm{F}-\nabla p)\varGamma_i (\bm u),
\end{equation}
where $\varGamma_i(\bm{u})=\omega_i +s_i(\bm{u})$. From Eq.~(\ref{conservedVariable}), the macroscopic quantities in every control volume are calculated by
\begin{gather}
\label{phi}   \phi  =\sum_i \tilde{g_i},\\
\label{u}
\bm u  =\frac{1}{c_s^2\rho}\left[\sum_i\bm{\xi}_i \tilde{f_i}+\frac{\Delta t}{2}c_s^2\bm F\right],\\
\label{P}
  p =\sum_i \tilde{f_i} +\frac{\Delta t}{2}c_s^2 \left(\bm{u}\cdot \nabla\rho -\gamma\rho\nabla\cdot(\lambda\nabla\mu_{\phi})\right).
\end{gather}

The kinetic viscosity $\nu$ and the mobility $\lambda$ are defined as, respectively,
\begin{equation}\label{kineticandmobility}
  \nu=c_s^2\tau_f, \hspace{6mm}           \lambda=c_s^2\tau_g\eta.
\end{equation}
Note that in present model the calculation formula of viscosity is different from that in the quasi-incompressible lattice Boltzmann model, i.g., $\nu=c_s^2 (\tau_f-0.5)$. In older to ensure the continuity of viscosity across the interface, the mixed dynamic viscosity  is given by \cite{zuhe2013}
\begin{equation}\label{mixvis}
  \mu=\frac{\mu_A \mu_B (\phi_A-\phi_B)}{(\phi-\phi_B)\mu_B+ (\phi_A-\phi)\mu_A},
\end{equation}
where $\mu_A=\rho_A \nu_A$, $\mu_B=\rho_B \nu_B$.
The first- and second-order derivatives can be approximated by different schemes~\cite{guo2011,yuan2006}. In this study, the first- and second-order derivatives are calculated as
\begin{equation}\label{nonphi}
\left. \nabla \Phi \right |_{\bm{x}}=\frac{1}{ab\delta_{\bm{x}} } (\theta_l^2 \Phi_{\bm{x}_{j+1}} + bc\Phi_{\bm{x}_{j}}- \theta^{2}_{r} \Phi_{\bm{x}_{j-1}} ),
\end{equation}
\begin{equation}\label{nonphi2}
\left. \nabla^2 \Phi \right |_{\bm{x}}=\frac{2}{ab\delta_{\bm{x}}^2} (\theta_l \Phi_{\bm{x}_{j+1}} -b\Phi_{\bm{x}_j}+ \theta_{r} \Phi_{\bm{x}_{j-1}} ),
\end{equation}
where $\bm x$ denotes the standard cartesian coordinates in two dimensions, $\theta_l=(\bm x _{j}-\bm x_{j-1})/\delta_{\bm{x}}$, $\theta_r=(\bm x_{j+1}-\bm x_{j})/\delta_{\bm{x}} $ are the forward and backward step lengths scaling factors, respectively, $\delta_{\bm{x}}$ is the grid size when $\theta_l=\theta_r$, and
 $a=\theta_l \theta_r$, $b=\theta_l+\theta_r$, $c=\theta_r- \theta_l$. For a uniform grid, the above formulas are equivalent to the central difference format with second-order accuracy. The detailed derivation process is shown in Appendix B.

 In summary, the  procedure in one time step of the present DUGKS is as follows:

 $\textbf{step 1.}$ Set the initial values of $\phi(\bm{x}_j,t_n)$, $\bm{u}(\bm{x}_j,t_n)$ and $p(\bm{x}_j,t_n)$, and compute the distribution functions $\varphi(\bm{x}_j,\bm{\xi}_i,t_n), \tilde{\varphi}(\bm{x}_j,\bm{\xi}_i,t_n)$ based on Eqs.~(\ref{feq}), (\ref{geq}) and (\ref{auxf}a) in each cell.

 $\textbf{step 2.}$ Compute  $\bar{\varphi}^+ (\bm{x}_j,\bm{\xi}_i,t_n)$ and $\tilde{\varphi}^+(\bm{x}_j,\bm{\xi}_i,t_n)$ according to Eqs.~(\ref{relations1}) and (\ref{relations2}) in each cell.

 $\textbf{step 3.}$ Compute $\bar{\varphi}^+(\bm{x}_b,\bm{\xi}_i,t_n)$ at the interface by linear interpolation, compute the $\bar{\varphi}(\bm{x}_b,\bm{\xi}_i,t_n+h)$ with Eq.~(\ref{fbar}).

 $\textbf{Step 4.}$ Compute the order parameter $\phi(\bm{x}_b,t_n+h)$, density $\rho(\bm{x}_b,t_n+h)$, velocity $\bm{u}(\bm{x}_b,t_n+h)$ and pressure $p$ at the interface from $\bar{\varphi}(\bm{x}_b,\bm{\xi}_i,t_n+h)$, then compute the original  distribution function $\varphi(\bm{x}_b,\bm{\xi}_i,t_n +h)$ with Eq.~(\ref{original-function}).

 $\textbf{Step 5.}$ Compute the microflux  across the cell interfaces from $\varphi(\bm{x}_b,\bm{\xi}_i,t_n +h)$   with Eq.~(\ref{microfluxf}).

 $\textbf{Step 6.}$ Update the distribution functions $\tilde{\varphi}(\bm{x}_j,\bm{\xi}_i,t_n+\Delta t)$  based on Eq.~(\ref{updatef}) in each cell.

 $\textbf{Step 7.}$ Update the values of $\phi(\bm{x}_j,t_n+\Delta t)$, $\rho(\bm{x}_j,t_n+\Delta t)$, $\bm{u}(\bm{x}_j,t_n+\Delta t)$ and $p(\bm{x}_j,t_n+\Delta t)$ via Eqs.~(\ref{oderrho}), (\ref{phi}-\ref{P}).

\section{Numerical Results and discussion}
In this section, several tests are performed to validate the accuracy and robustness of the proposed DUGKS method, including a two-dimensional stationary droplet, a layered Poiseuille flow and a bubble rising problem.
In each test case comparisons with the existing LBE models are also performed.
In all simulations, $RT$ is fixed at $1/3$ and  $C_{max}$ is set to be $\sqrt{6RT}$ unless otherwise stated.

\subsection{A stationary droplet}
\begin{figure}
\centering
\includegraphics[width=0.75\textwidth]{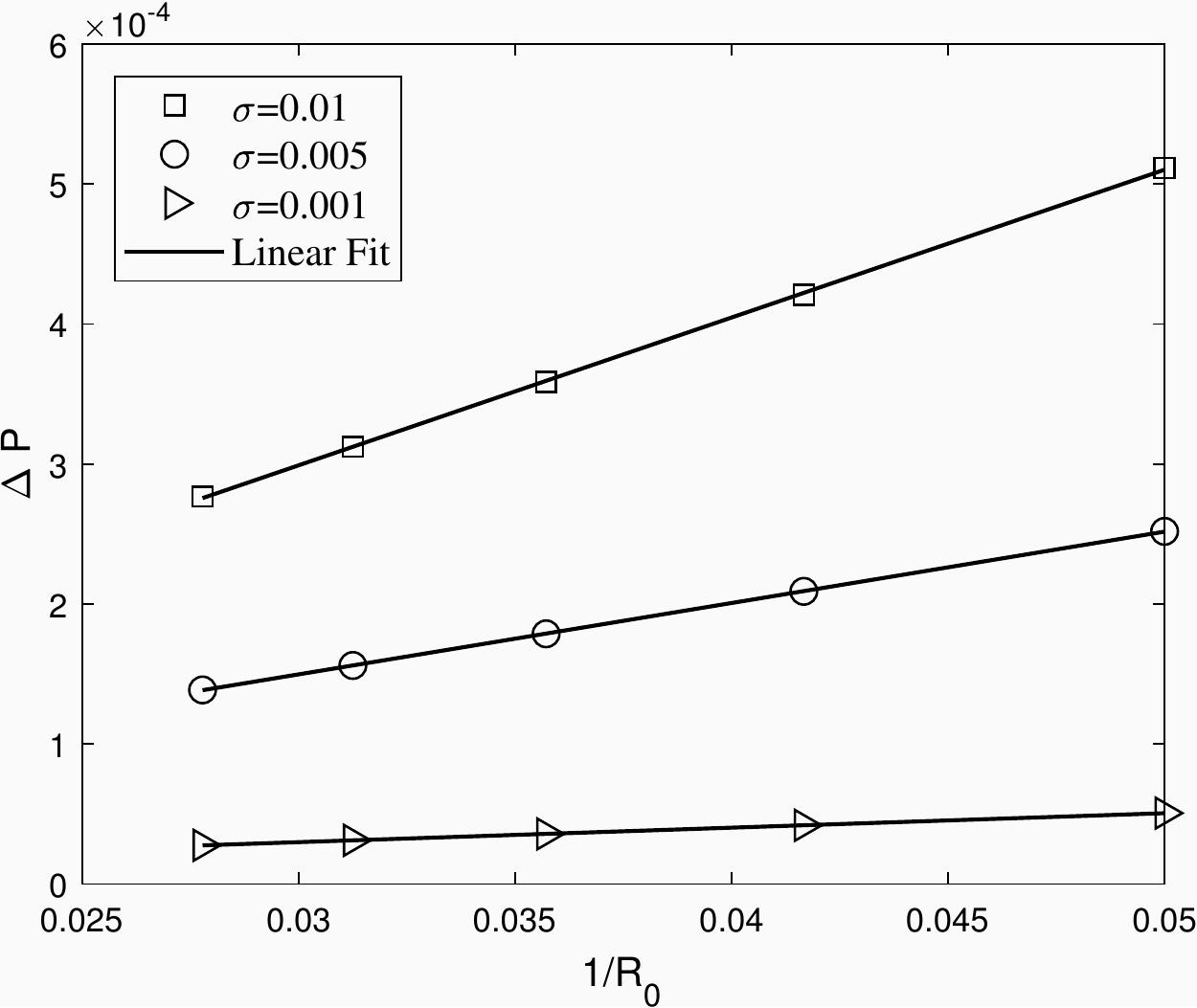}
\caption{Test of Laplace's law with $\sigma=0.01,0.005$ and  $0.001$.}\label{laplace}
\end{figure}

\begin{figure}
\centering
\includegraphics[width=0.75\textwidth]{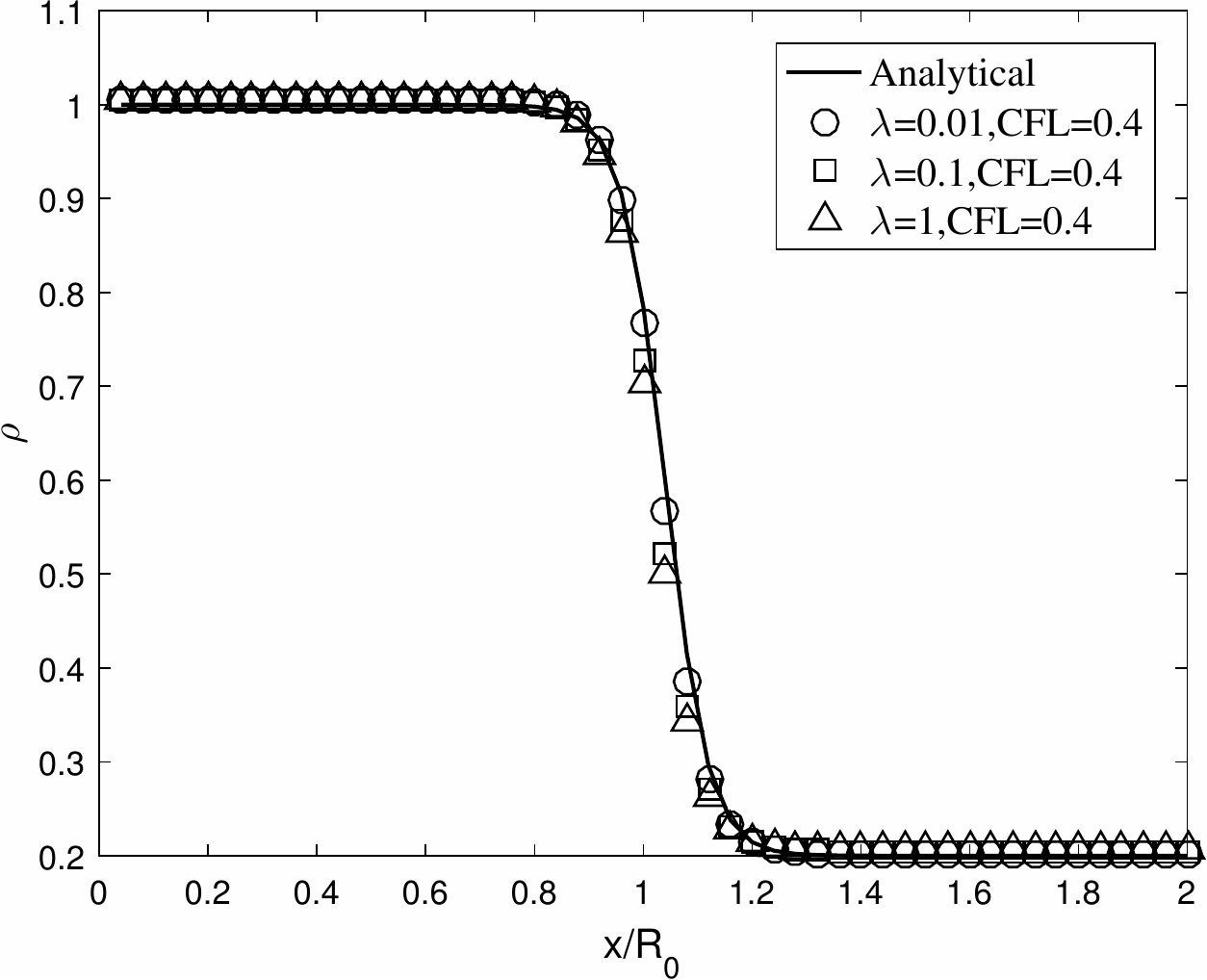}
\caption{Density profile across the interface with different  $\lambda $ and $\mbox{CFL}=0.4$.}\label{figA1}
\end{figure}

 The first test  is a stationary droplet immersed in another fluid. This problem is used to assess the capability of the proposed model in handling the surface force. Initially, a circular droplet with radius ranging from $10$ to $40$ (in lattice unit) is placed at the center of a square computational domain with periodic boundary conditions at all boundaries. The domain is divided into $100\times 100$ uniform cells. The order parameter is initialized as
\begin{equation}\label{droplet}
\phi(x,y)=\frac{\phi_A+\phi_B}{2}+\frac{\phi_A-\phi_B}{2}\times\tanh\left(\frac{2( R_0-\sqrt{(x-x_c)^2+(y-y_c)^2})}{W}\right),
\end{equation}
where $(x_c,y_c)$ is the center position of the computational domain, $R_0$ is the droplet radius. The model parameters are given by $\rho_A=1$, $\rho_B=0.2$,  $\tau_f=\tau_g=0.5$, $ \phi_A=1$, $\phi_B=0$, $W=4$, $\mbox{CFL}=0.4$ and  $\sigma=0.001$.
First,  we will test the Laplace's law. When the equilibrium state is reached, the pressure distribution across the interface will be proportional to the inverse of the radius, i.e., $\Delta P=\sigma /R_0 $, where $P$ is obtained by
$P=p_0-\kappa\phi\nabla^2\phi+ \kappa|\nabla\phi |^2/2+ p$ with the equation of state $ p_0=\phi\partial_{\phi}{\psi}-\psi $ \cite{lee2009,zuhe2013}. Therefore, the surface tension can be calculated by $\sigma=R_0 \Delta P $.
 Figure.~\ref{laplace} shows the relation between the pressure difference and the reciprocal of the radius. According to the Laplace law, the surface tensions from by our model are agree well with the theoretical values.
 The density profiles with three values of mobility are shown in Fig.~\ref{figA1} as a function of the radial distance from the center of the droplet normalized by $R_0$. We can observe that the density profiles agree well with the analytical shape. However, a slight deviation at the interface grows as the value of mobility increasing. The same situation also exists in Refs ~\cite{zuhe2013,liang2014}. This is because the total energy can be reduced by shrinking the drop by shifting the bulk $\phi$ slightly away from the initial values. As a result, it is not conserved for the enclosed mass of the droplet calculated by the median level of the order parameter~\cite{liyibao2016}.

\begin{figure}
\centering
\includegraphics[width=0.75\textwidth]{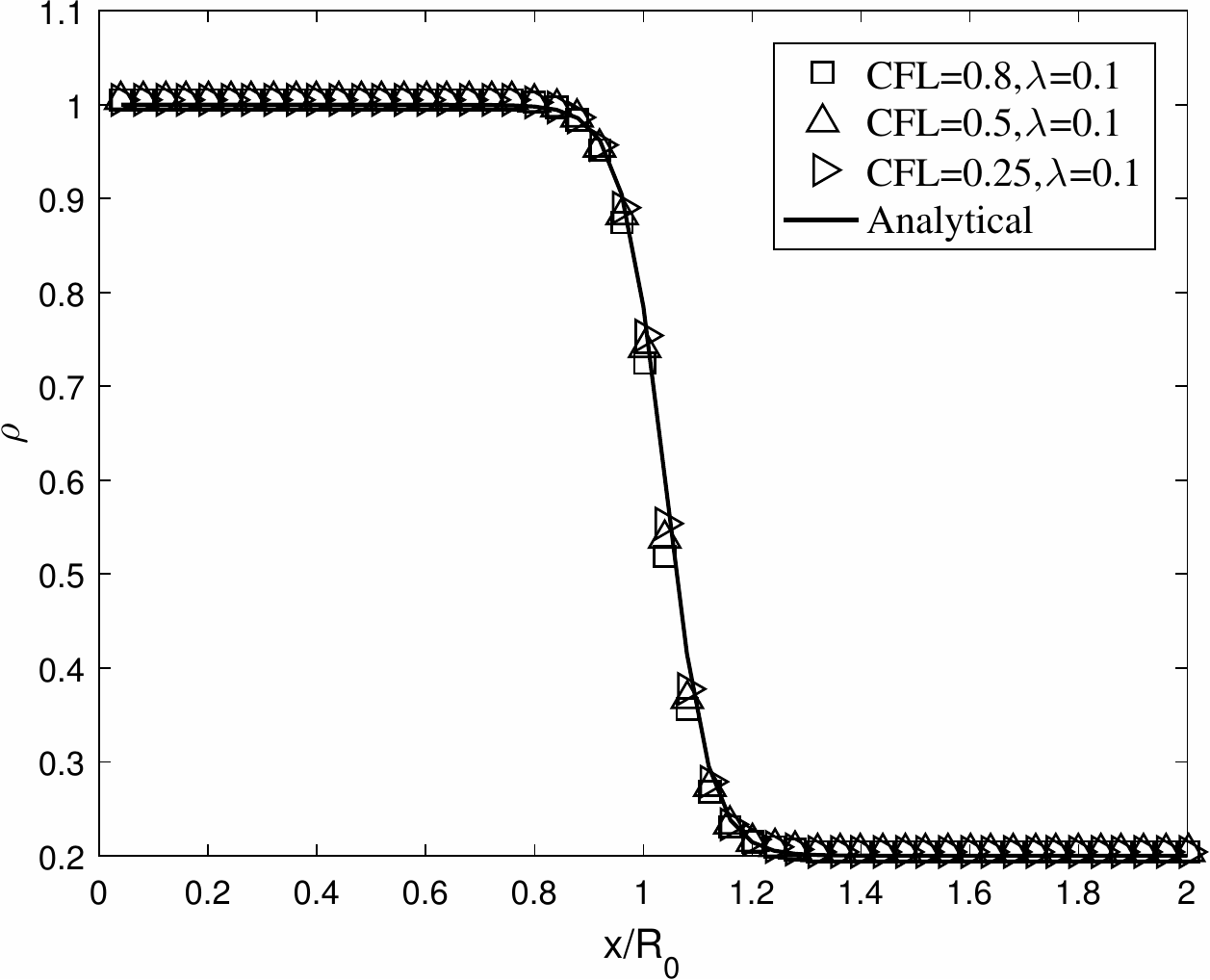}
\caption{Density profile across the interface with different $\mbox{CFL}$ numbers.}\label{caseAcfl}
\end{figure}

Now we investigate the effects of the CFL number by repeating the above test with a fixed $\lambda=0.1$. Since the grid size and the sound speed of the flow are fixed, changing the CFL number actually changes the time step. The results are shown in Fig.~\ref{caseAcfl}. It can be observed the density profiles obtained with different CFL numbers agree well with the theoretical values. And the results with a smaller CFL number (or time step) agree better with the analytical ones.

Although both the quasi-incompressible LBE model~\cite{yang2016} and the present model can recover the mass conservation equation through the Chapman-Enskog analysis, the present model can improve the mass conservation property due to the numerical scheme. To illustrate this point, we compare the equilibrium mass to the original total mass of the droplet with different radius to evaluate local mass conservation property. The results are shown in Table~\ref{tableA1}, in which the relative error is defined as
$(M_{ini}-M_{ter})/M_{ini}\times 100\%$, where $M_{ini}$ and $M_{ter}$ are the initial and final steady masses of the droplet, respectively. It can be observed  that the mass loss increases as the radius of the droplet decreasing for both models, however the present model can keep the mass conservation more accurate than the LBE model~\cite{yang2016}.

\begin{table}[htbp]
\centering
\caption{\label{tableA1} Relative mass errors of the present model and LBE model \cite{yang2016} for a circular droplet.}
\begin{tabular}{p{2cm} p{2cm} p{2cm} p{2cm} p{2cm} p{2cm}}
\hline\hline
$R_0$ & $M_{ini}$ &Ref.\cite{yang2016}  & Error(\%) &Present &Error(\%)\\
\hline
  36 &4049\quad &4025\quad & 0.59 \quad &4029\quad &0.49\quad \\
  32\quad &3205\quad &3149\quad & 1.75 \quad &3173\quad &0.99\quad \\
  28\quad &2449\quad &2409\quad & 1.63 \quad &2421\quad &1.14\quad\\
  24\quad &1789\quad &1749\quad & 2.24 \quad &1763 \quad&1.45\quad \\
  20\quad &1245\quad &1185\quad & 4.82 \quad &1209 \quad&2.89\quad \\
\hline\hline
\end{tabular}
\end{table}

\subsection{Layered Poiseuille flow}
\begin{figure}
\centering
\begin{tabular}{cc}
\includegraphics[width=0.49\textwidth]{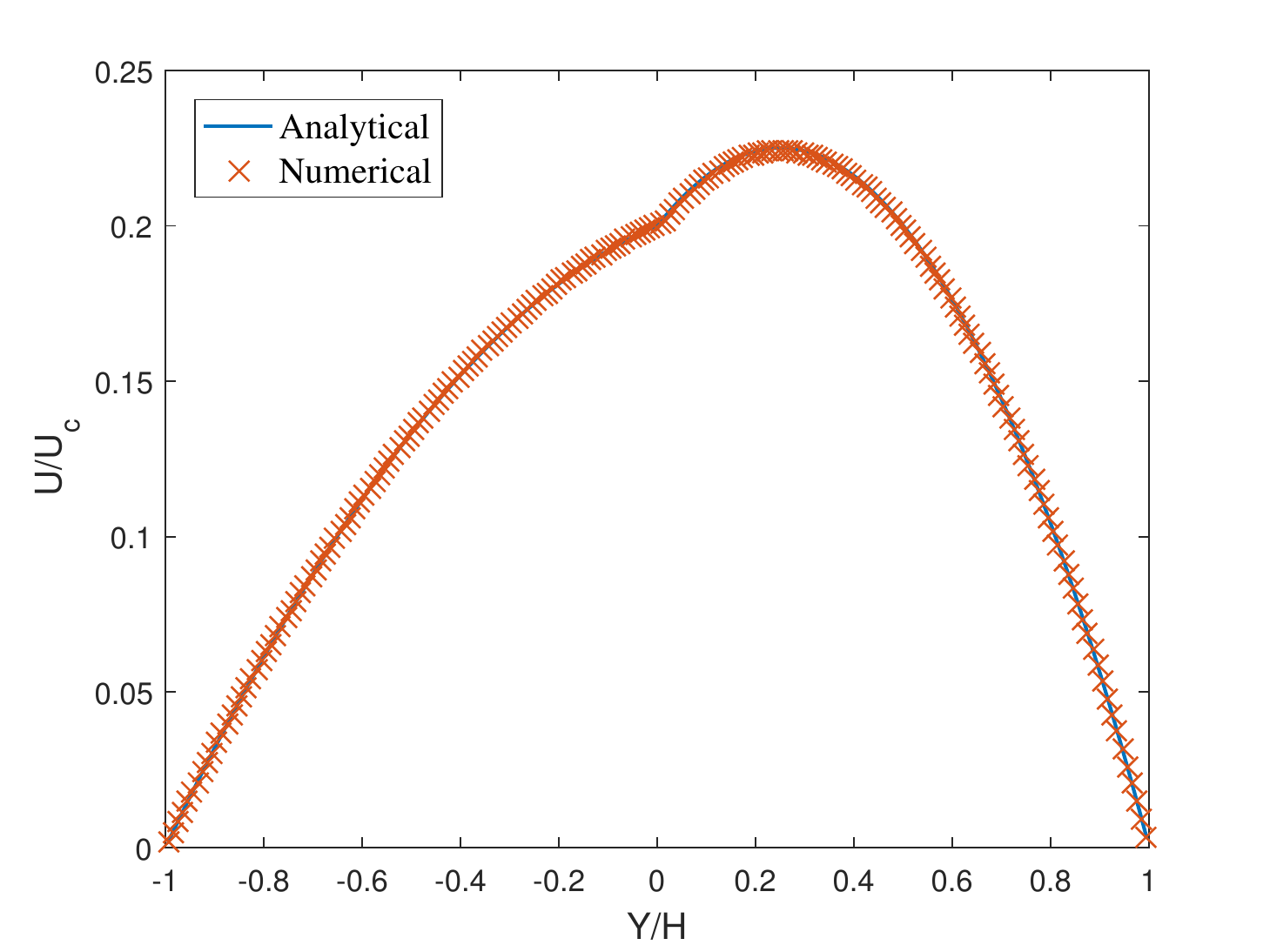}&
\includegraphics[width=0.49\textwidth]{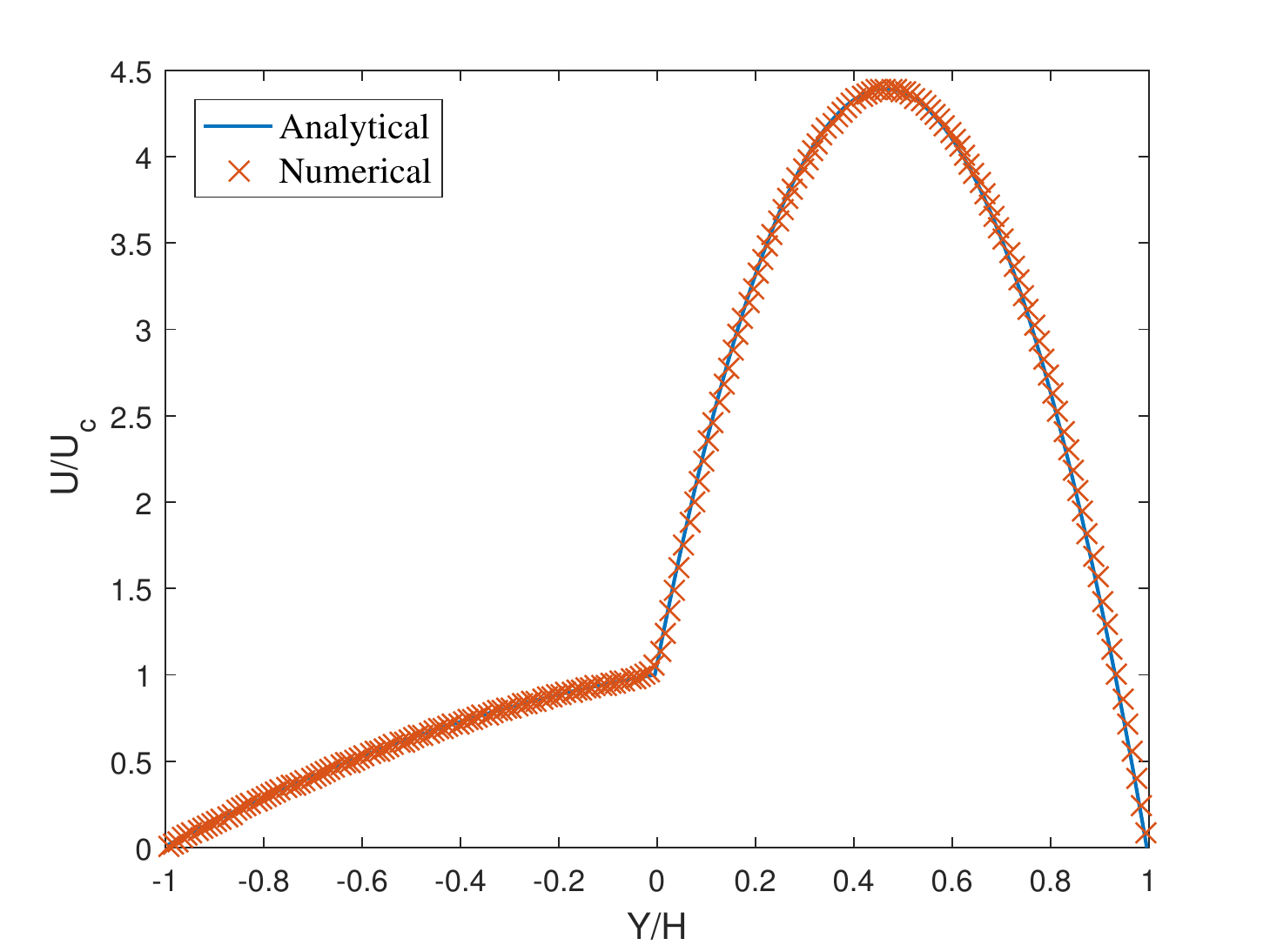}\\
(a)&(b)\\
\includegraphics[width=0.49\textwidth]{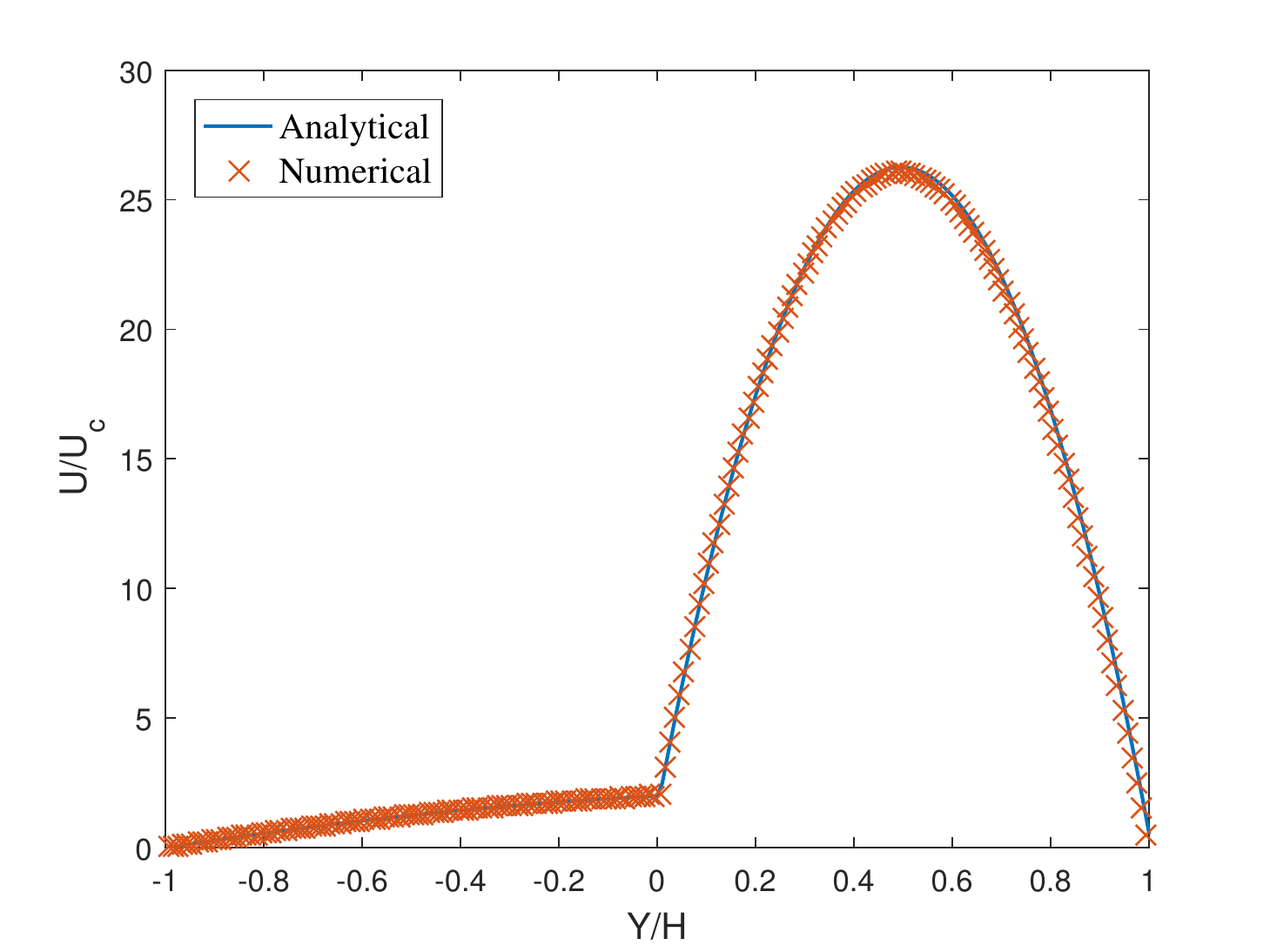}&
\includegraphics[width=0.49\textwidth]{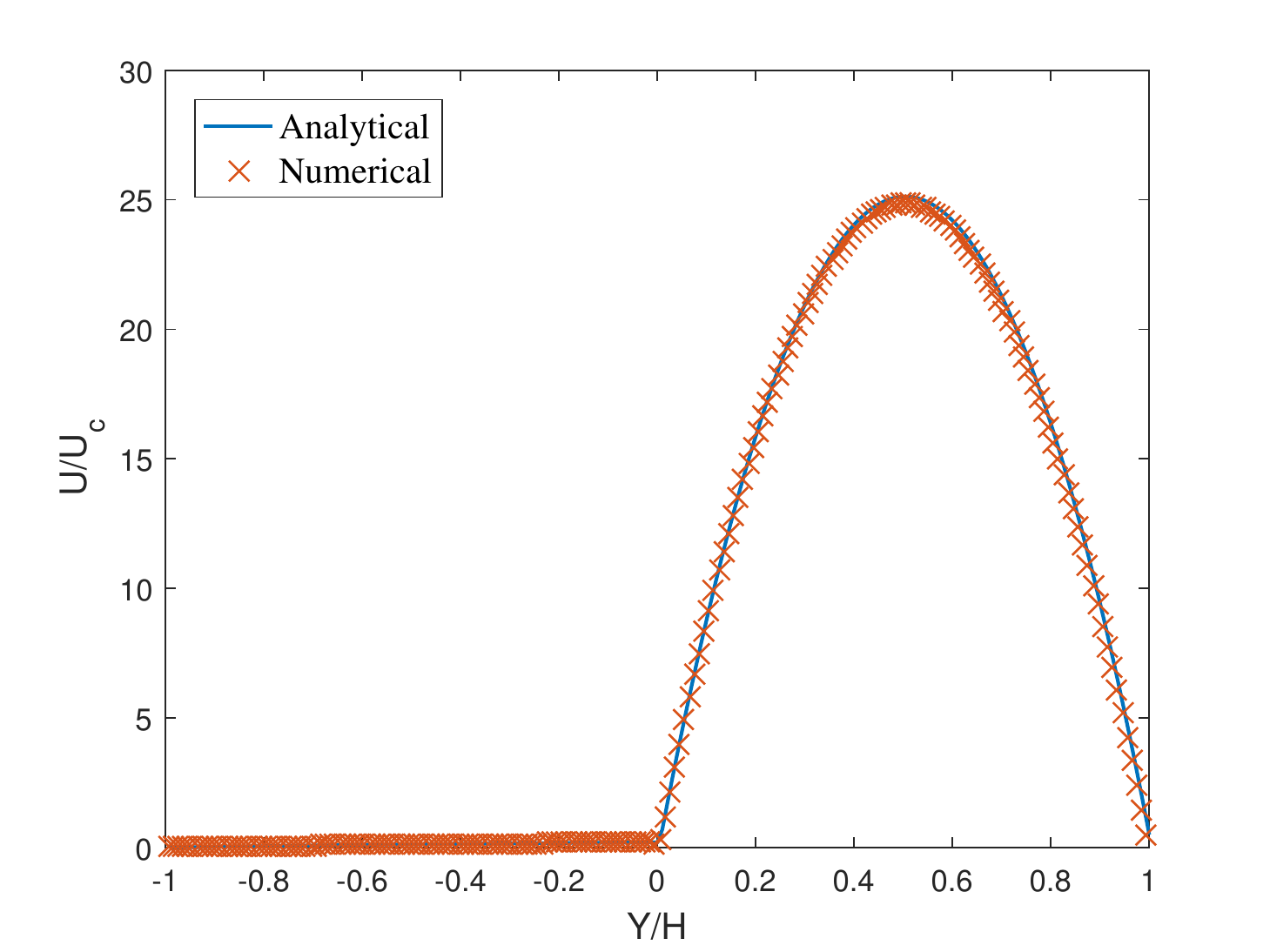}\\
(c)&(d)\\
\end{tabular}
\caption{Comparison of the velocity profile of layered Poiseuille flow obtained by the present method with the analytical results. (a) $\mu_A/\mu_B=3$; (b) $\mu_A/\mu_B=30$; (c) $\mu_A/\mu_B=100$; (d) $\mu_A/\mu_B=1000.$}
\label{layervelocity}
\end{figure}

To validate the capability of the present DUGKS for simulating binary fluids with different viscosities, a layered Poiseuille flow of two immiscible fluids (denoted by A and B) between two infinite plates located at $y=H$ and $-H$ is now simulated.
In the test, fluid $A$ is filled in the region $0<y\leq H$ while fluid B is filled in the region $-H\leq y<0$, and the channel width is $2H$.
The flow is driven by a pressure gradient $G$ in the flowing direction. When the flow is sufficiently slow and no instabilities occur at the interface, an analytical solution with a steady velocity profile exists,
\begin{equation}
\bm u_{x,a}(y)=
\begin{cases}
\frac{GH^2}{2\mu_A} \left[-(\frac{y}{H})^2- \frac{y}{H}( \frac{\mu_A-\mu_B}{\mu_A+\mu_B}) + \frac{2\mu_A}{\mu_A+ \mu_B}\right], & \mbox{$0\leq y\leq H$} \\
\frac{GH^2}{2\mu_B} \left[-(\frac{y}{H})^2- \frac{y}{H}( \frac{\mu_A-\mu_B}{\mu_A+\mu_B}) + \frac{2\mu_B}{\mu_A+\mu_B}\right], & \mbox{$-H\leq y\leq 0$}
\end{cases}
\end{equation}
The steady velocity at the center can be determined once a pressure gradient $G$ is given, i.e., $u_c=G H^2/(\mu_A+\mu_B)$.
In the simulation, a uniform mesh of $10 \times 200$ is used.
Periodic boundary conditions are applied to the inlet and outlet of the channel, and no-slip boundary conditions are enforced on the two walls. The steady velocity at the center is set to be $5\times10^{-5}$ to ensure the stability of the interface. It is worth pointing out that artificially adding body force to mimic the pressure gradient is not precisely valid in the presence of a density contrast~\cite{zuhe2013}. Thus, a binary fluid with the same density is considered here. Four different viscosity ratios of $\mu_A/\mu_B=3,30,100,1000$ are considered in the simulations. Other parameters are set as $W=4, \rho_A=\rho_B=1$ and $\mbox{CFL}=0.5$.
 Velocity profiles are normalized by the central velocity and shown in Fig.~\ref{layervelocity}.
 As shown in Fig.~\ref{layervelocity}, the predicted velocity profiles agree well with the analytical solutions in all cases considered.

\begin{figure}
\centering
\includegraphics[width=0.5\textwidth]{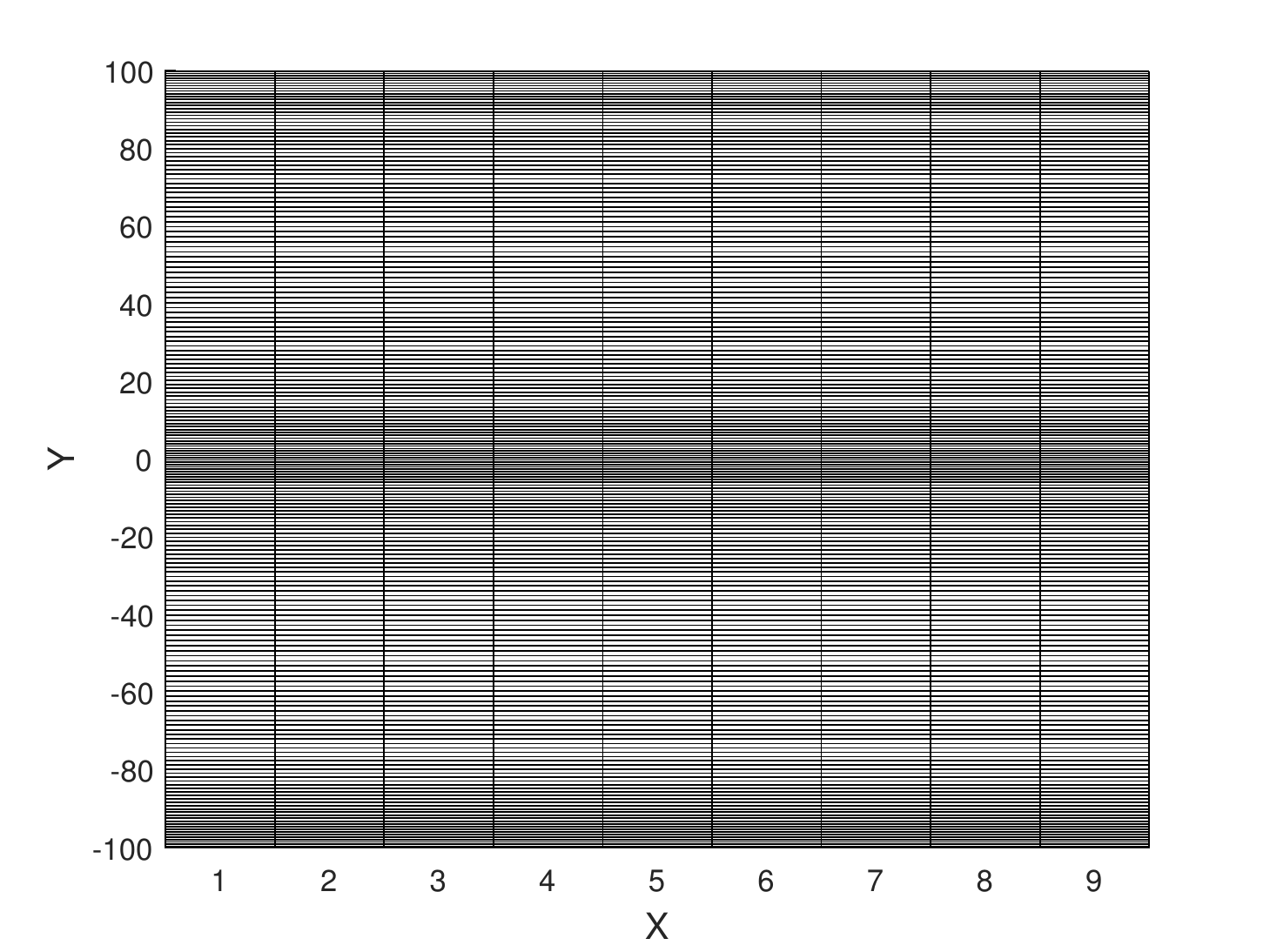}
\caption{Schematic of the nonuniform grid for layered Poiseuille flow with $\epsilon=2.5$.}\label{nonuniform}
\end{figure}

\begin{figure}
\centering
\includegraphics[width=0.5\textwidth]{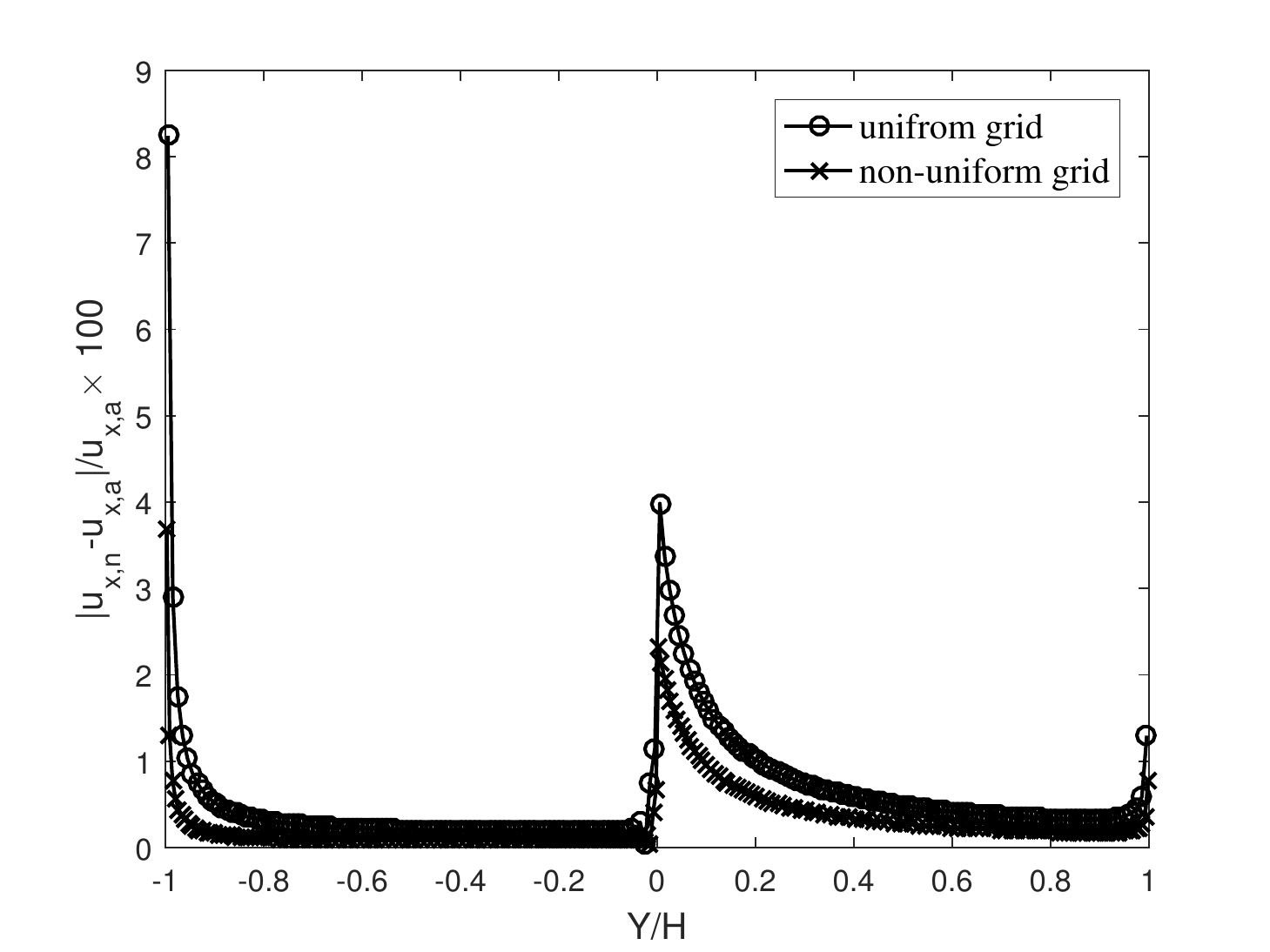}
\caption{Relative errors for layered Poiseuille flow at the dynamic viscosities ratio $\mu_A/\mu_B=30$.}\label{layerror}
\end{figure}

To improve the predictions, we repeat the above simulation by using a locally refined mesh in the vicinities of the walls and phase interface, as shown in Fig.~\ref{nonuniform}. The coordinates in the $y$ direction are generated by $y_i/H=( \xi_i+\xi_{i+1})/2$ for $-H\le i\le H$, where $\xi_i$ is defined by
\begin{equation}
\xi_i=
\begin{cases}
\frac{1}{2}+\frac{\tanh(\epsilon(i/H-0.5) )}{ 2\tanh(\epsilon/2)},\hspace{2mm}  & \mbox{$0\leq i\leq H$} \\
-\frac{1}{2}+\frac{\tanh(\epsilon(i/H+0.5) )}{ 2\tanh(\epsilon/2)},\hspace{2mm} & \mbox{$-H\leq i\leq 0$},
\end{cases}
\end{equation}
where $\epsilon$ is an adjustment coefficient that determines the distribution of the grid. Generally, a larger value of $\epsilon$ leads to a finer mesh near the endpoints. In the present test, $\epsilon$ is set to be 2.5. To be clear, the velocity relative errors with $\mu_A/\mu_B=30$ are shown in Fig.~\ref{layerror}, where the relative error is defined as the absolute value of the discrepancy between the numerical velocity $u_{x,n}$ and analytical velocity $u_{x,a}$ divided by the analytical solution $u_{x,a}$. From Fig.~\ref{layerror}, it is clear that the relative errors drop significantly in comparison with those using the uniform mesh, particularly near the transition region and the wall.

\subsection{Rising bubble}
In this section, a bubble rising due to buoyancy is used to test the capability of the present DUGKS for simulating binary fluids with different densities.
A light circular bubble (fluid A) with diameter $D=2L_0/5$ is immersed in another fluid (fluid B) with higher density. Initially, the bubble is positioned at
$(L_0/2,L_0/2)$ in a rectangular domain of size $L_0\times3L_0$. In the simulations, periodic boundary conditions are applied to all boundaries. The buoyancy force $F_{b,y}=-(\rho-\rho_B)G_y$, where $G_y$  is the magnitude of the gravitational acceleration in the $y$ direction, is applied to the fluids.
The dynamic behavior of a rising bubble mainly involves five dimensionless parameters, namely, the ratios of density and viscosity of the two fluids, the Eotvos (or Bond) number, the Morton number, and the Reynolds number, which are defined as~\cite{takada2001}
\begin{equation}\label{dimensionless}
  \mbox{Eo}=\frac{G_y(\rho_B-\rho_A)D^2}{\sigma },    \mbox{Mo}=\frac{G_y(\rho_B-\rho_A)\mu_B^4}{\rho_B^2\sigma^3},  \mbox{Re}=\frac{\sqrt{G_y\rho_B(\rho_B-\rho_A)D^3}}{\mu_B}.
\end{equation}
The bubble shape depends on these non-dimensional parameters under different flow regimes~\cite{clift1978,hua2007}.

\begin{figure}
\centering
\begin{tabular}{cc}
\includegraphics[width=0.49\textwidth]{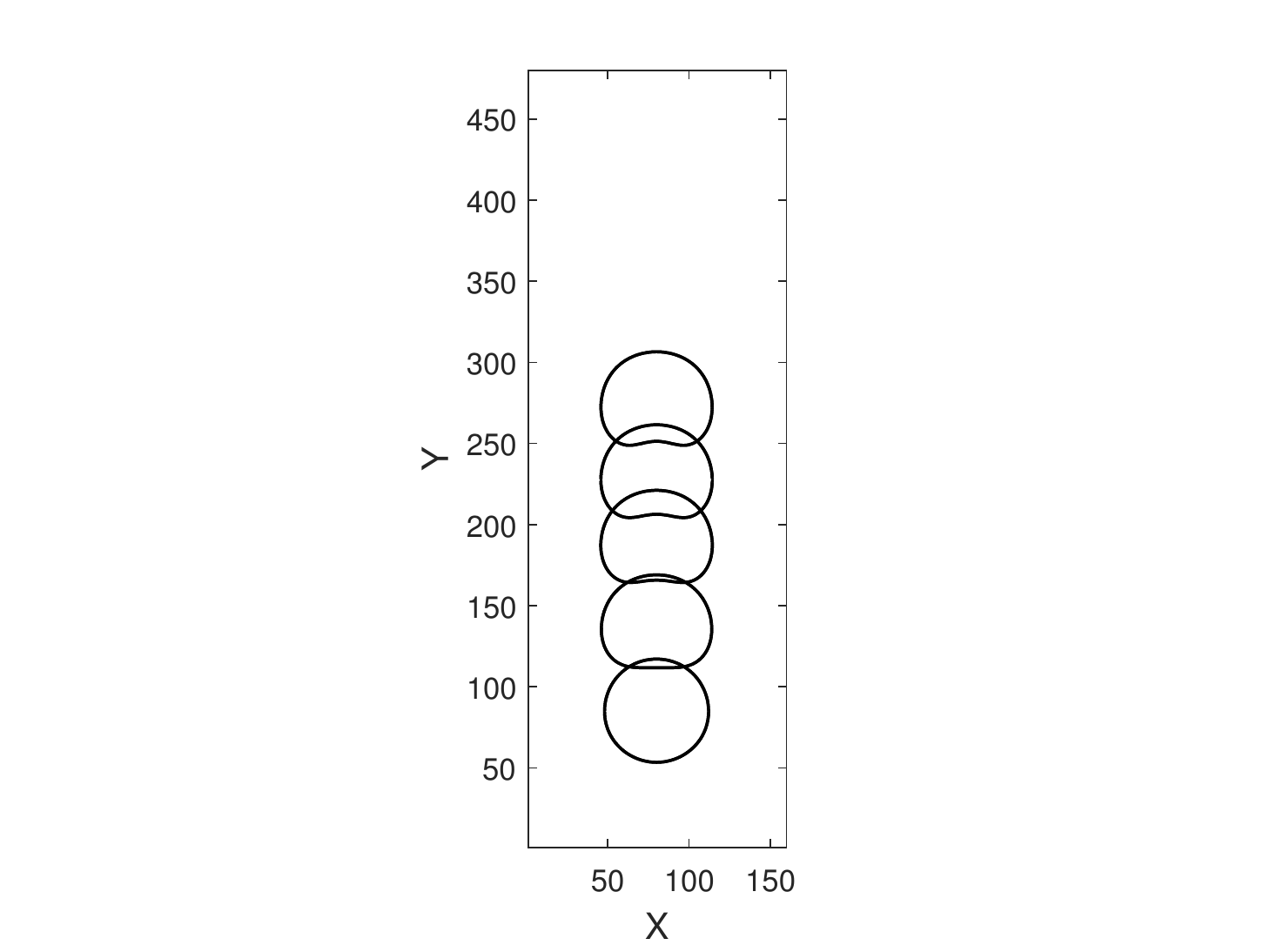} &
\includegraphics[width=0.49\textwidth]{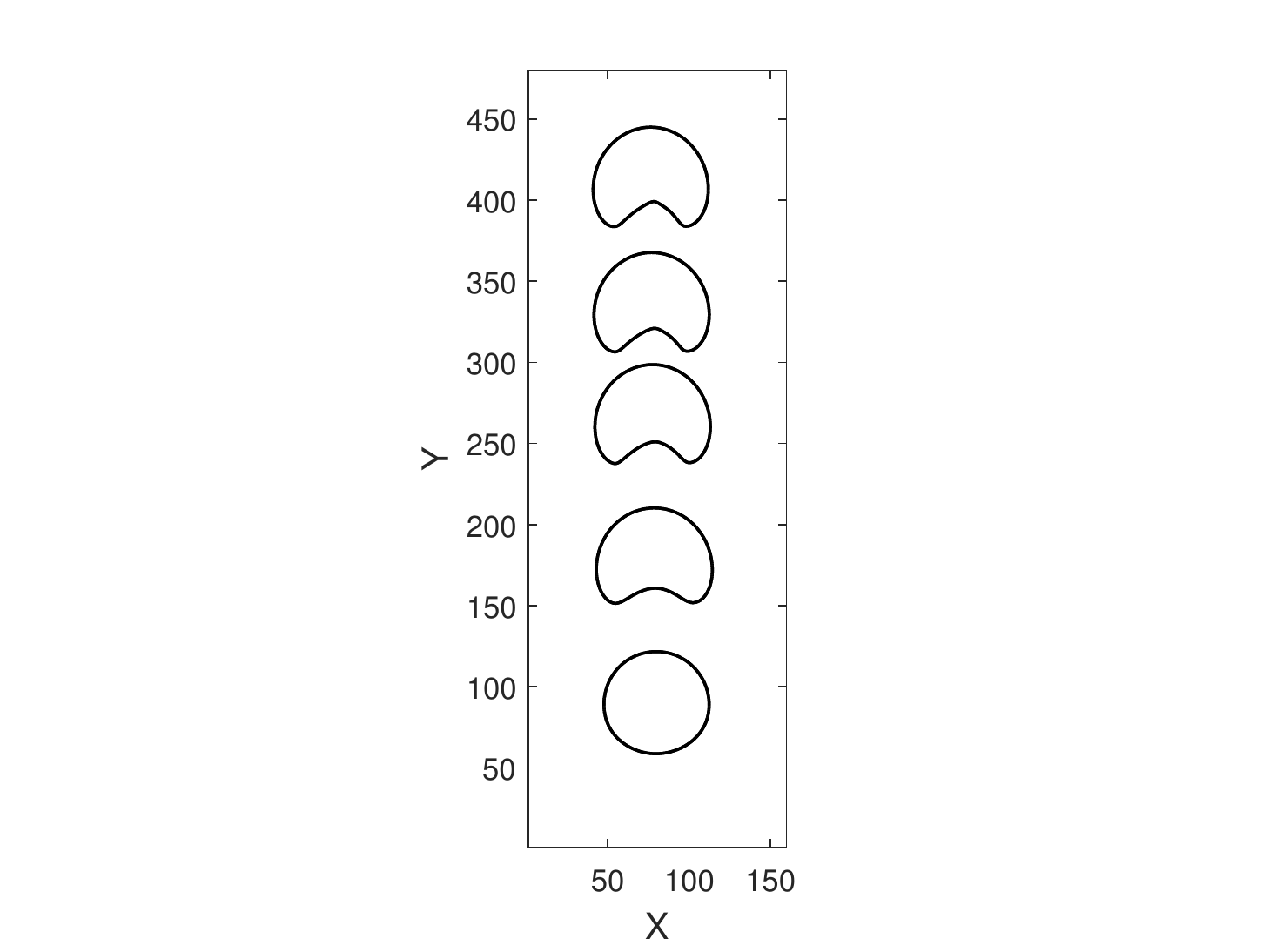}
\end{tabular}
\caption{Evolution of the rising bubble at $t^*=0.988, 4.94, 7.91, 9.88, 11.86$. (a) $\rho_B/\rho_A=2$ ($\mbox{Eo}=20.48,\mbox{Mo}=3.86,\mbox{Re}=6.87$), (b) $\rho_B/\rho_A=5$ ($\mbox{Eo}=39.32,\mbox{Mo}=10.67,\mbox{Re}=8.69$).}
\label{caseC1}
\end{figure}

In order to compare the results with the LBE model in Ref.~\cite{yang2016},
 the parameters in the simulations are set to be  $\ G_y=10^{-5}$, $\sigma=0.001$, $L_0=160$, $\eta=2.0$, $\mbox{CFL}=0.354$ and
 $W=4$. The viscosity ratio is set to be unity to stay compatible with the model in Ref.~\cite{yang2016}.
Figure~\ref{caseC1} shows the evolution of the interface shape obtained by the present method at different dimensionless times which are defined by $t^*=t\sqrt{G_y/D}$.
From Fig.~\ref{caseC1} (a), it is seen that the results of the model in Ref.~\cite{yang2016} and the present model are nearly identical when $\rho_B/ \rho_A=2$. However, for a higher density ratio, e.g., $\rho_B/ \rho_A=5$, the LBE model becomes unstable while the present model can still give satisfactory predictions.
  Figure~\ref{caseC1} (b) shows the evolution of the interface shape obtained by the present model with $\rho_B/ \rho_A=5$.

\subsection{ Rayleigh-Taylor instability}
To further demonstrate the capacity of the present model in solving more complicated flows, we conducted a simulation of the Rayleigh-Taylor instability (RTI) at high Reynolds numbers, which occurs when a slight perturbation at the interface between a heavy fluid  and a light one arises in a gravitational field. This is a classical problem that has been extensively studied by experimental measurements~\cite{waddell2001} and
numerical methods~\cite{hechen1999, liang2014}.

\begin{figure}
\centering
\begin{tabular}{c}
\includegraphics[width=1\textwidth,trim=0 70 0 40,clip]{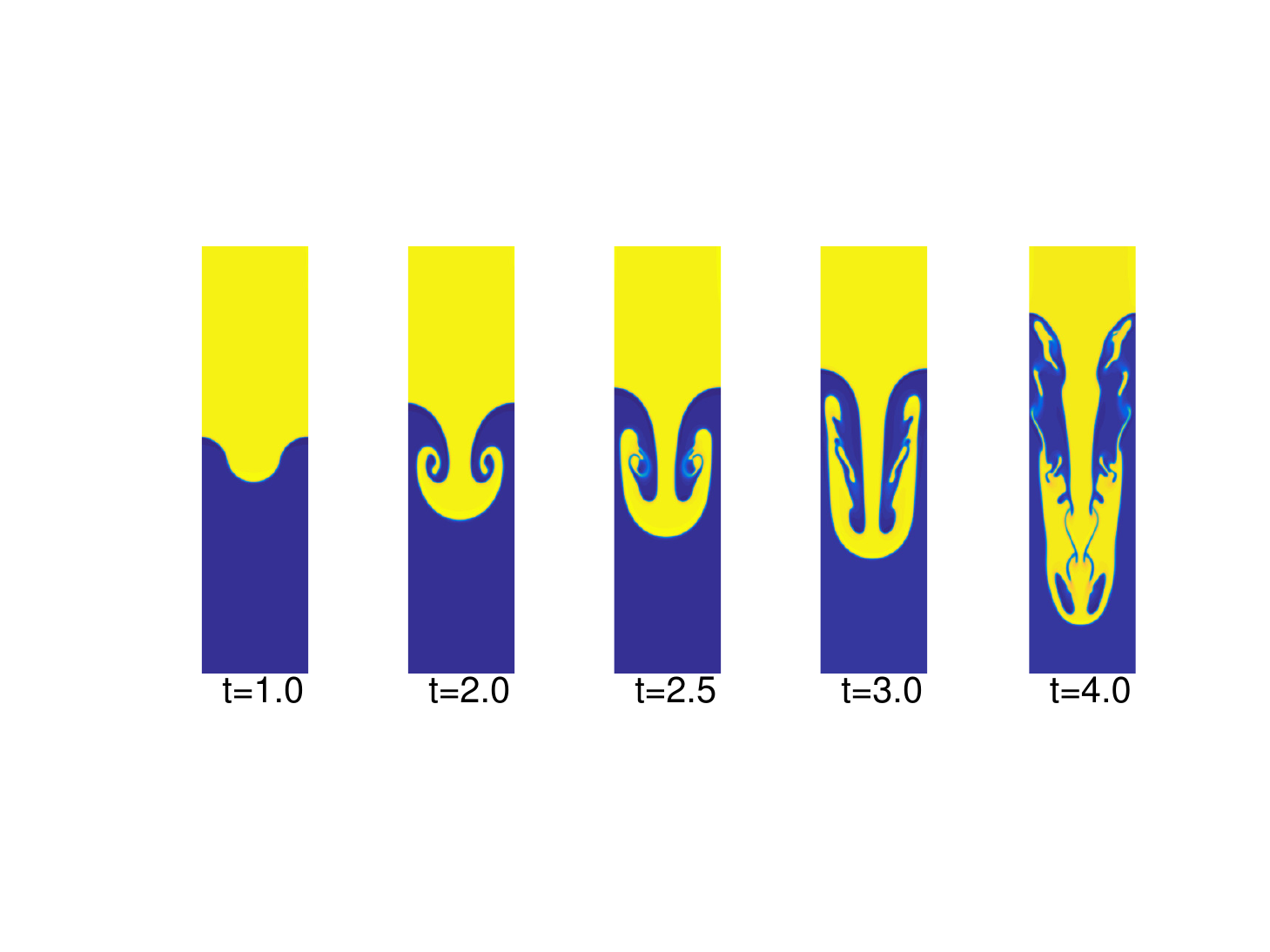} \\
(a)\\
\includegraphics[width=1\textwidth,trim=0 70 0 40,clip]{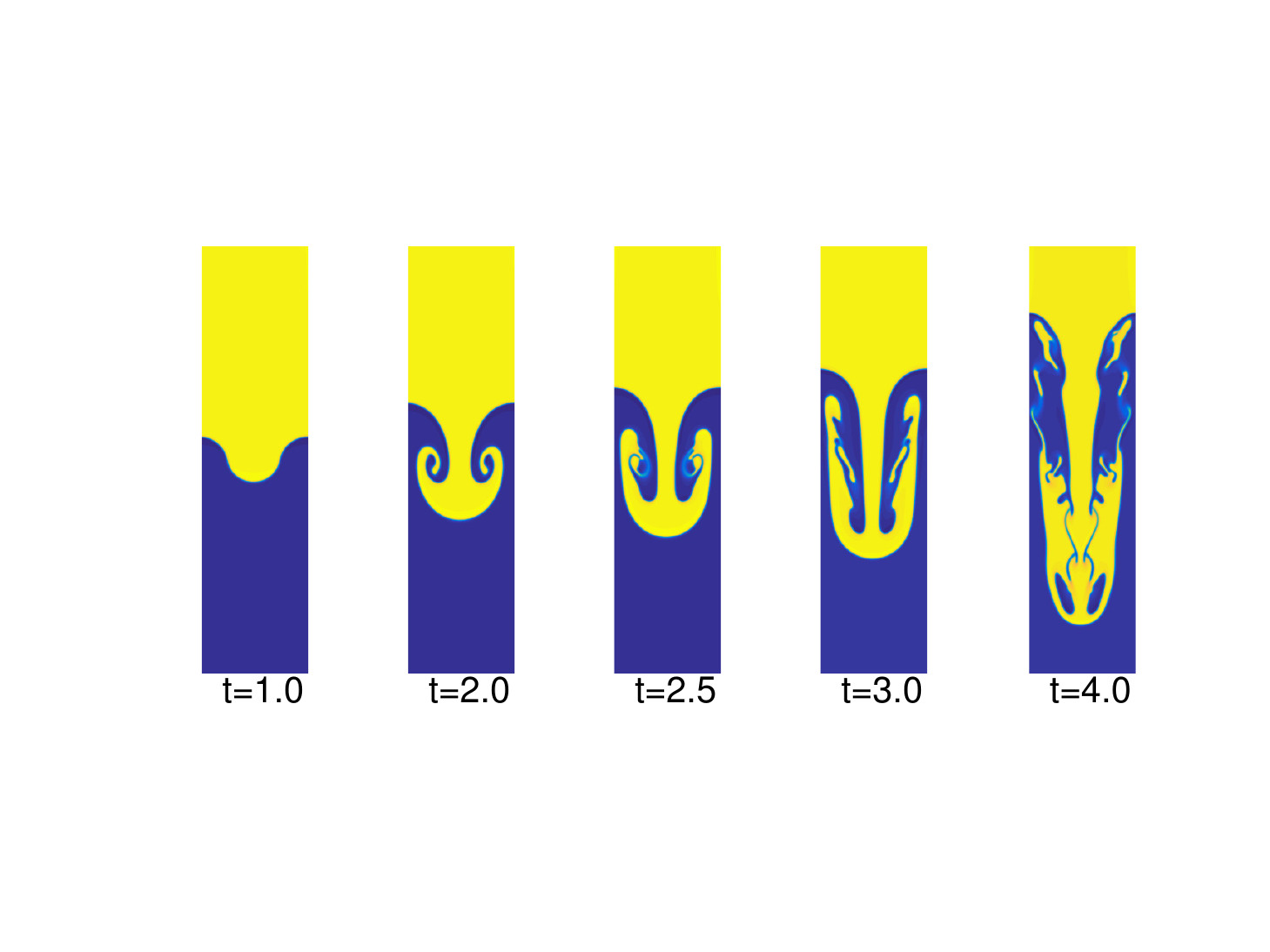}\\
(b)
\end{tabular}
\caption{Evolution of the interface patterns of the Rayleigh-Taylor instability at (a) $\mbox{A}_t=0.1$, $\mbox{Re}=150$, (b) $\mbox{A}_t=0.1$, $\mbox{Re}=3000$. The time is normalized by the characteristic time $T=\sqrt{L/Ag}/\Delta t$.}
\label{caseD1}
\end{figure}

\begin{figure}[ht]
\centering
\includegraphics[width=0.9\textwidth,trim=0 70 0 40,clip]{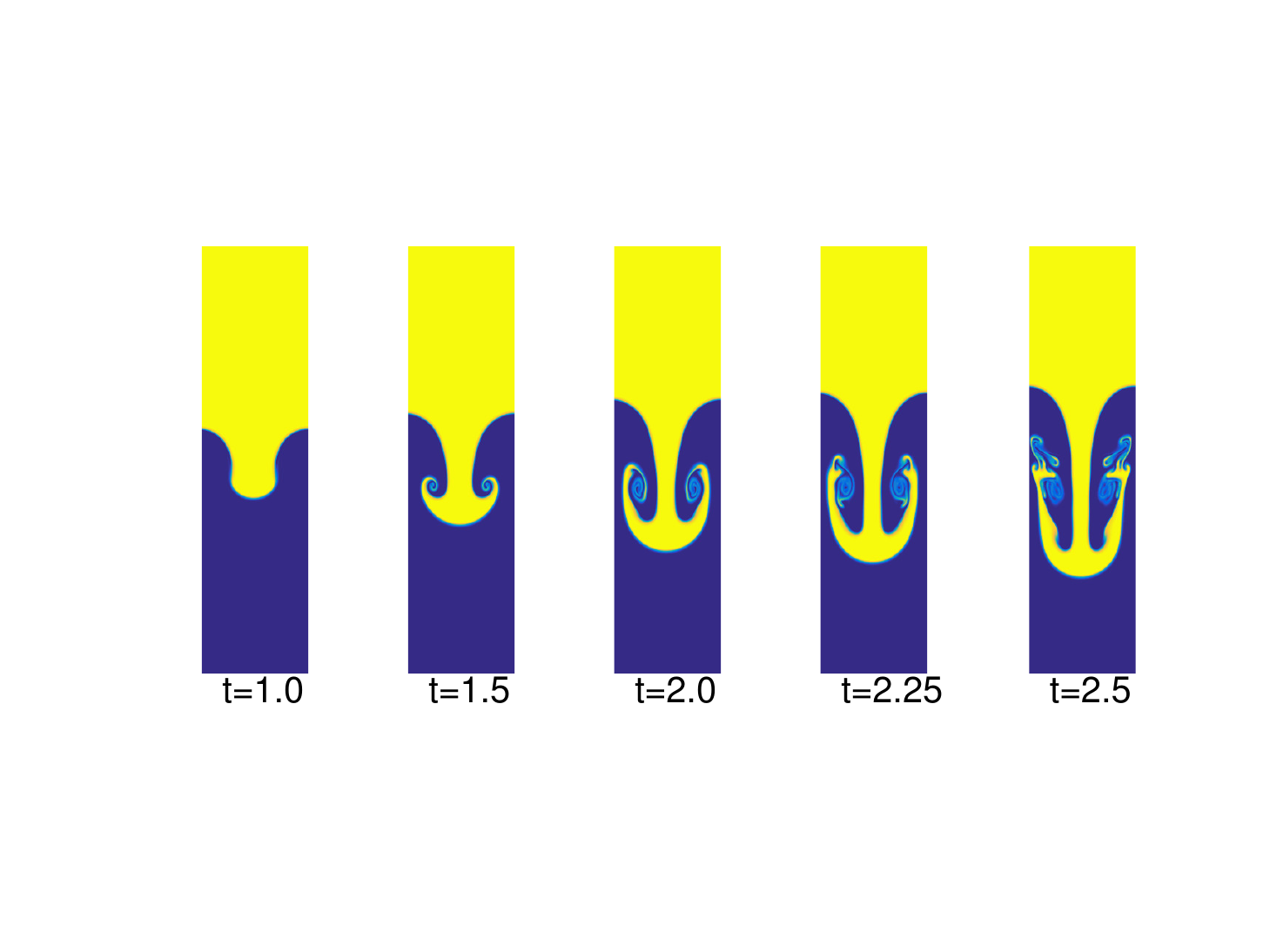}
\caption{Evolution of the interface patterns of the Rayleigh-Taylor instability at $\mbox{A}_t=0.5$ and $\mbox{Re}=3000$.}
\label{fig9}
\end{figure}

\begin{figure}
\centering
\begin{tabular}{cc}
\includegraphics[width=0.49\textwidth]{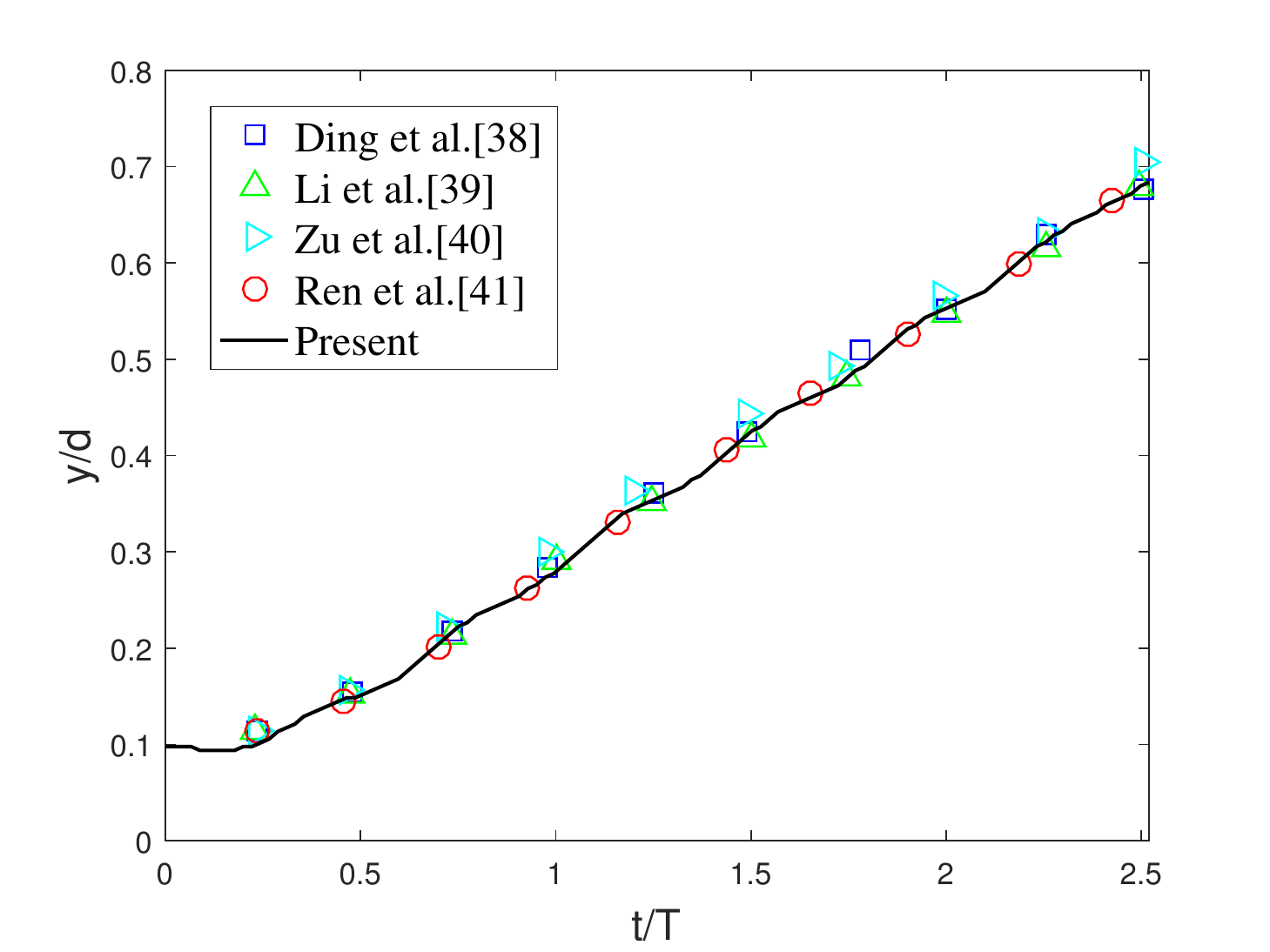}&
\includegraphics[width=0.49\textwidth]{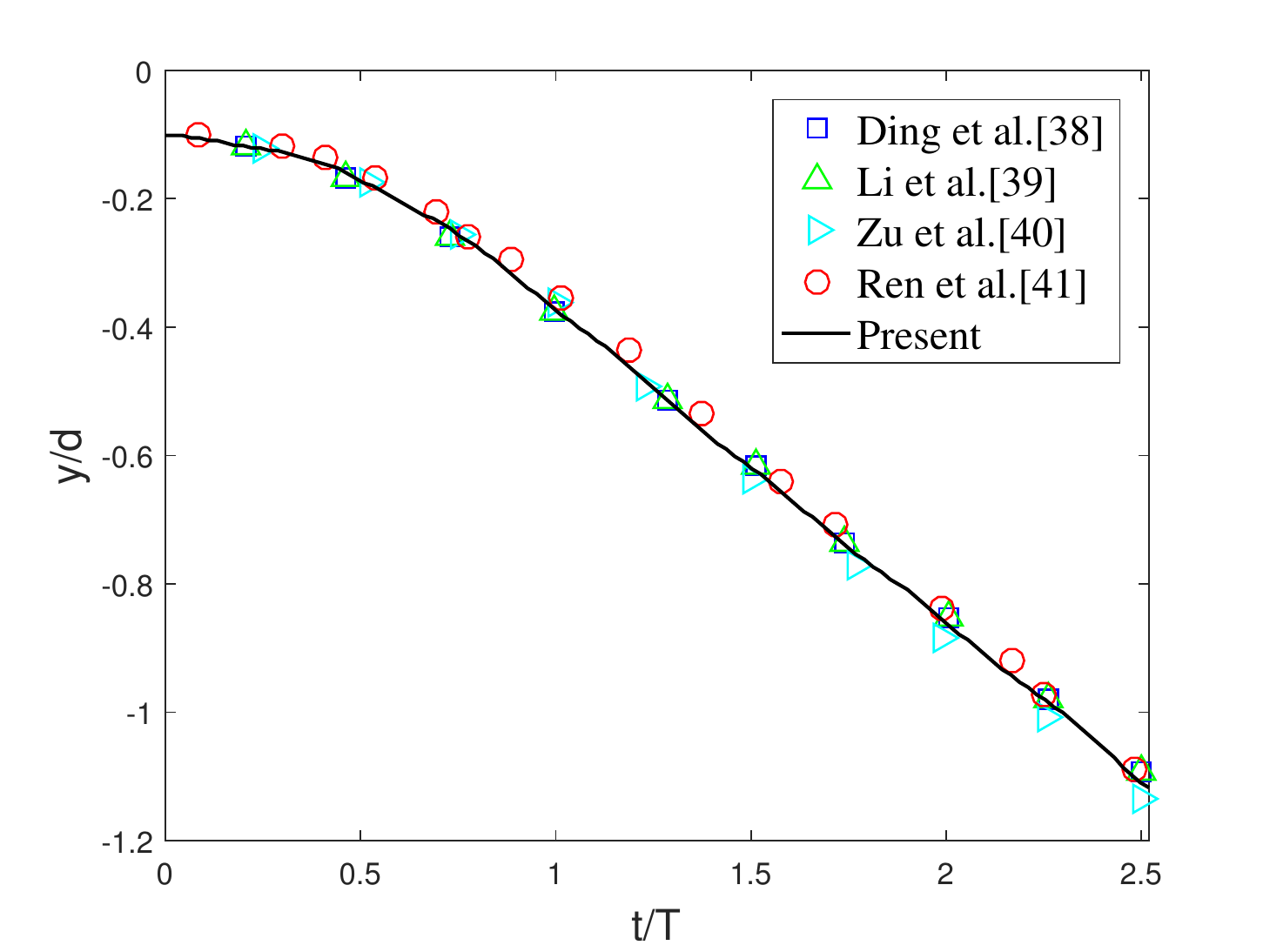}\\
(a)&(b)\\
\end{tabular}
\caption{Time evolution of the positions of (a) the bubble front and (b) the spike tip, and comparisons with the results of Ding $\emph{et al.}$ ~\cite{ding2007}, Li $\emph{et al.}$~\cite{Li2012} Zu $\emph{et al.}$~\cite{zuhe2013} Ren $\emph{et al.}$~\cite{renf2016}.}
\label{caseD2}
\end{figure}

In the simulation we set the Atwood number $\text{A}_t=(\rho_A-\rho_B)/(\rho_A+\rho_B)=0.1$ and  Reynolds number $\mbox{Re}=\rho_A d^{3/2}g^{1/2}/\mu=3000$, where $g$ is the gravitational acceleration pointing downward. The computational domain is $[0,d]\times[-2d,2d]$ and the initial interface of the two fluids is $H(x,y)=2d+0.05d\cos(2\pi x/L)$, where $L$ is the wavelength. The bounce-back boundary conditions are applied to the bottom and top boundaries and periodic boundary conditions are imposed on the lateral boundaries. The other parameters are set as $d=L=256, \sqrt{gL}=0.04,  \mbox{W}=4, \mbox{CFL}=0.35 $ and $\sigma=5.0\times 10^{-5}$.
These parameters are the same as used in the work of Liang $\emph{et al.}$~\cite{liang2014} except for the CFL.
The evolution of the interface at dimensionless times $ t=1T,\ 2T,\ 2.5T,\ 3T,\ 4T$ is shown in Fig.~\ref{caseD1}, where $T$ is the characteristic time defined as $T=\sqrt{L/{Ag}}/\Delta t$.
 It can be seen that the interfacial patterns agree well with those reported in Ref.~\cite{liang2014}. In addition, for further comparison with previous literature results, a test with $\text{A}_t=0.5$, $\text{Re}=3000$ and $\text{CFL}=0.283$ is also simulated and shown in Fig.~\ref{fig9}.
 The quantitative comparison of the time histories  of the  bubble front and spike tip  is shown in Fig.~\ref{caseD2}, which shows an excellent agreement between the results from other studies~\cite{ding2007,Li2012,zuhe2013,renf2016}.

\section{CONCLUSIONS}
In this study, we extend the DUGKS method to two-phase flows based on a quasi-incompressible phase-field theory which can exactly remain the  mass conservation. With the finite volume scheme, better accuracy is expected. To validate the performance of the proposed model, a series of numerical test are performed.

First, with the two-dimensional stationary droplet test, it is demonstrated that the proposed DUGKS model satisfies Laplace's law and the adjustable time step can improve the numerical accuracy.
Furthermore, the tests of the layer Poiseuille flow with large viscosity radios and the bubble rising with higher density radios demonstrate the superior numerical stability compared with the LBE model. In particular, by adopting a non-uniform mesh, the present model can reduce numerical errors near the interface and the fixed boundary. In order to further illustrate the model capability of dealing with complicated interface, the Rayleigh-Taylor instability phenomenon is also successfully simulated. The validity and capacity of the present model are well demonstrated.

\section*{ACKNOWLEDGEMENTS}
This study was supported by the National Key Research and Development Plan (Grant No. 2016YFB0600805).

\begin{appendix}
\section{CHAPMAN-ENSKOG ANALYSIS}
\label{derivation}
In this section, the present DUGKS model for hydrodynamic equations are analyzed through the Chapman-Enskog expansion. We first expand the hydrodynamic distribution function with the time and space derivatives in consecutive scales of $\epsilon$, which keeps the same order of magnitude of the Knudsen number,
\begin{align}\label{fexpend}
f_i=f^{(0)}_i+\epsilon f^{(1)}_i+\epsilon^2 f^{(2)}_i+\ldots,
\end{align}
\begin{equation}\label{texpend}
\partial_t =\epsilon\partial_{t_0}  + \epsilon^2 \partial_{t_1}, \hspace{5mm}\nabla=\epsilon \nabla_0,\hspace{5mm}F_i=\epsilon F^{(0)}_i+ \epsilon^2 F^{(1)}_i,
\end{equation}
with
\begin{equation}\label{Fexpend}
\begin{split}
& F_i^{(0)}= (\bm{\xi}_i-\bm{u})\cdot [ \Gamma_i(\bm{u}) \bm{F} +  s_i c_s^2 \nabla \rho],\\
& F_i^{(1)}=-\omega_i c_s^2 \rho \gamma\nabla \cdot (\lambda \nabla \mu_{\phi}).
\end{split}
\end{equation}
By substituting these into the Eq (\ref{bgkf}) and equalling the equation with respect to  the same order of $\epsilon$, we have
\begin{equation}\label{f0infinite}
O(\epsilon^0)\mbox{:} \hspace{5mm} f_i^{(0)}=f_i^{eq}
\end{equation}
\begin{equation}\label{f1infinite}
O(\epsilon^1) \mbox{:} \hspace{5mm} \partial_{t_0} f_i^{(0)} + \bm{\xi}_i \cdot \nabla_0 f_i^{(0)} = - \frac{1}{\tau_f} f_i^{(1)} + F_i^{(0)}
\end{equation}
\begin{equation}\label{f2infinite}
O(\epsilon^2) \mbox{:} \hspace{5mm} \partial_{t_0} f_i^{(1)} + \bm{\xi}_i \cdot \nabla_0 f_i^{(1)} + \partial_{t_1} f_i^{(0)}= - \frac{1}{\tau_f} f_i^{(2)} + F_i^{(1)}
\end{equation}
From the definitions Eq~(\ref{feq}) and (\ref{F}), we have
\begin{equation}\label{sum f cf}
\sum_i{f_i^{eq}}=p, \hspace{2mm} \sum_i \xi_i f_i^{eq}= c_s^2 \rho \bm{u}, \hspace{2mm} \sum_i \xi_i \xi_i f_i^{eq}=c_s^2 p+c_s^2\rho\bm{u} \bm{u}
\end{equation}
\begin{equation}\label{sum cccf}
\sum_i{\xi_i \xi_i \xi_i f_i^{eq}}=c_s^4\rho ( u_\alpha\delta_{\beta\gamma}+ u_\beta \delta_{\alpha\gamma} + u_\gamma \delta_{\alpha\beta}),
\end{equation}
\begin{equation}\label{sumF}
\sum_i F_i^{(0)}=c_s^2 \bm{u}\cdot\nabla \rho, \hspace{2mm} \sum_i F_i^{(1)}=- c_s^2 \rho\gamma \nabla \cdot (\lambda \nabla\mu_{\phi}),
\end{equation}
\begin{equation}\label{sumcF}
\sum_i \xi_i F_i^{(0)}=c_s^2 \bm{F}, \hspace{2mm} \sum_i \xi_i F_i^{(1)}=0.
\end{equation}
\begin{equation}\label{sumccF}
\sum_i \xi_i \xi_i F_i^{(0)}=c_s^2 (\bm{F'u+uF'})+ c_s^4\bm{u}\cdot \nabla \rho, \hspace{2mm} \sum_i \xi_i \xi_i F_i^{(1)}= c_s^4\rho\gamma \nabla \cdot (\lambda \nabla \mu_{\phi}) .
\end{equation}
Then, taking the zeroth- to second-order moments of Eq. (\ref{f1infinite}) gives
\begin{equation}\label{f1m0}
\partial_{t_0} p + \nabla_0 \cdot ( c_s^2 \rho \bm{u}) =c_s^2 \bm{u} \cdot \nabla \rho,
\end{equation}
\begin{equation}\label{f1m1}
\partial_{t_0} (c_s^2\rho\bm{u}) + \nabla_0 \cdot ( c_s^2 p + c_s^2 \rho \bm{uu}) =c_s^2 \bm{F},
\end{equation}
\begin{equation}\label{f1m2}
\begin{split}
\partial_{t_0} (c_s^2 p+c_s^2\rho\bm{u}\bm{u}) + \nabla_0 \cdot c_s^4 \rho ( u_{\alpha} \delta_{\beta\gamma}+ u_{\beta} \delta_{\alpha \gamma} +u_{\gamma} \delta_{\alpha\beta})=\\
-\frac{1}{\tau_f} \sum \xi_i \xi_i f_i^{(1)} +c_s^2 (\bm{F'u+uF'}) +c_s^4 \bm{u} \cdot \nabla\rho.
\end{split}
\end{equation}
Likewise, taking the zeroth- and first-order moments of Eq. (\ref{f2infinite}) gives
\begin{equation}\label{f2m0}
\partial_{t_1} p= -c_s^2 \rho \gamma \nabla \cdot (\lambda \nabla \mu_{\phi}),
\end{equation}
\begin{equation}\label{f2m1}
\partial_{t_1} (c_s^2\rho\bm{u})+\nabla_0 \cdot \sum{\bm{\xi \xi} f_i^{(1)}}= 0.
\end{equation}
The Eq. (\ref{f1m0}) can be rewritten as
\begin{equation}\label{f1m0new}
\partial_{t_0} p + c_s^2 \rho \nabla_0 \cdot (\bm{u})=0,
\end{equation}
Combining Eqs. (\ref{f2m0}) and (\ref{f1m0new}) leads to
\begin{equation}\label{continuity-equation}
\frac{1}{c_s^2\rho} \partial_t p + \nabla \cdot \bm{u}= -\gamma \nabla \cdot (\lambda \nabla\mu_{\phi}).
\end{equation}
According to Eqs.  (\ref{f1infinite}),  (\ref{f2m0}) and (\ref{f1m0new}), the second-order moment of $f_i^{(1)}$ in Eq. (\ref{f2m1}) becomes
\begin{equation} \label{ccf1}
\sum \xi_i \xi_i f_i^{(1)}= -\tau_f c_s^4 \rho ( \nabla \bm{u} + \nabla \bm{u}^T)+O(Ma^3).
\end{equation}
Substituting  Eq. (\ref{ccf1}) into Eq. (\ref{f2m1}) gives
\begin{equation}\label{moment-1}
\partial_{t_1} ( c_s^2 \rho \bm{u}) + \nabla_0 \cdot ( -\tau_f c_s^4 \rho ( \nabla \bm{u} + \nabla \bm{u}^T))=0.
\end{equation}
Combining Eqs. (\ref{f1m1}) and (\ref{moment-1}) leads to
\begin{equation}\label{momentequation}
\partial_t (\rho \bm{u}) + \nabla \cdot ( p+ \rho \bm{uu})= \nabla \cdot \rho \nu ( \nabla \bm{u} +\bm{u}^T) +\bm{F},
\end{equation}
where $ \nu=\tau_f  c_s^2$ is the kinematic viscosity.

Next, the CH equation will be derived based on Eq.~(\ref{bgkg}) through the Chapman-Enskog expansion. Similarly, the order distribution function is expanded as
\begin{equation}\label{gexpend}
g_i=g^{(0)}_i+\epsilon g^{(1)}_i+\epsilon^2 g^{(2)}_i+\ldots,
\end{equation}
\begin{equation}\label{nabla}
\partial_t =\epsilon \partial_{t_0}  + \epsilon^2 \partial_{t_1}, \hspace{5mm}\nabla=\epsilon \nabla_0,\hspace{5mm}G_i=\epsilon G^{(0)}_i,
\end{equation}
By substituting these into the Eq (\ref{bgkg}) and equalling the equation with respect to  the same order of $\epsilon$, we have
\begin{equation}\label{g0}
O(\epsilon^0) \mbox{:} \hspace{5mm} g_i^{(0)}=g_i^{eq}
\end{equation}
\begin{equation}\label{g1}
O(\epsilon^1)\mbox{:} \hspace{5mm} \partial_{t_0} g_i^{(0)} + \bm{\xi}_i \cdot \nabla_0 g_i^{(0)} = - \frac{1}{\tau_g} g_i^{(1)} + G_i^{(0)}
\end{equation}
\begin{equation}\label{g2}
O(\epsilon^2)\mbox{:} \hspace{5mm} \partial_{t_0} g_i^{(1)} + \partial_{t_1} g_i^{(0)}+ \bm{\xi}_i \cdot \nabla_0 g_i^{(1)} = - \frac{1}{\tau_g} g_i^{(2)}.
\end{equation}
From the definitions Eq.~(\ref{geq}) and (\ref{G}), we have
\begin{equation}\label{sum g cg}
\sum_i{g_i^{eq}}=\phi, \hspace{2mm} \sum_i \xi_i g_i^{eq}= \phi \bm{u}, \hspace{2mm} \sum_i \xi_i \xi_i g_i^{eq}=c_s^2\eta\mu_{\phi} + \phi \bm{uu},
\end{equation}
\begin{equation}\label{sumG}
\sum_i G_i^{(0)}=0,  \hspace{2mm} \sum_i \xi_i G_i^{(0)}= \frac{\phi}{\rho}\bm{G},
\end{equation}
where $\bm{G}=\bm{F}-\nabla p$. Then, taking the zeroth- and first-order moments of Eqs. (\ref{g1}) and (\ref{g2}) gives
\begin{equation}\label{g1-0-moment}
\partial_{t_0} \phi + \nabla_0 \cdot (\phi \bm{u}) =0,
\end{equation}
\begin{equation}\label{g1-1-moment}
\partial_{t_0} \phi \bm u+ \nabla_0 (c_s^2 \eta \mu_{\phi} +\phi \bm{uu})= -\sum \frac{1}{ \tau_g} \bm{\xi}_i g_i^{(1)} +\frac{\phi}{\rho}\bm{G},
\end{equation}
\begin{equation}\label{g2-0-moment}
\partial_{t_1} \phi + \sum_i \xi_i \cdot \nabla_0 g_i^{(1)}=0.
\end{equation}
And, substituting Eq. (\ref{g1-1-moment}) into (\ref{g2-0-moment}) leads to
\begin{equation}\label{g2-0-momentnew}
\partial_{t_1} \phi - \tau_g \nabla_0\cdot [\partial_{t_0} (\phi\bm{u})+ \nabla_0 \cdot (\phi \bm{uu}+c_s^2\eta \mu)-\frac{\phi}{\rho}\bm{G} ]=0.
\end{equation}
Assembling Eqs. (\ref{oderrho}) and (\ref{f1m1}), Eq. (\ref{g2-0-momentnew}) can be reduced to
\begin{equation}\label{g-t1}
\partial_{t_1}\phi=\nabla^2 (\lambda\mu_{\phi}),
\end{equation}
where $\lambda=c_s^2 \eta \tau_g$ is the mobility coefficient. Combining Eqs. (\ref{g1-0-moment}) and (\ref{g-t1}) leads to
\begin{equation}\label{ch-equation}
\partial_t \phi + \nabla \cdot (\phi\bm{u})= \lambda \nabla^2\mu_{\phi}.
\end{equation}

\section{A finite difference scheme for non-uniform grids}
Consider a one-dimensional computed region $ x\subseteq [a_0,a_1]$.
Without loss of generality, divide $[a_0,a_1]$ into N sub-intervals, not necessarily of equal length, by the points
$ a_0=x_0,x_1,x_2,\ldots,x_{N-1},x_{N}=a_1$. The forward and backward lengths scaling factors are marked by
$\theta_l=(x_i-x_{i-1})/\delta_{x} $,$\theta_r=(x_{i+1}-x_{i})/\delta_{x} $ with $\delta_{x}=1/N$. For a sufficiently smooth function $\Phi$, derivatives at interior grid points $x_i$, can be expended by Taylor's theorem as
\begin{equation}\label{forward}
\Phi_{i+1}=\Phi_i+ \theta_r \delta_{x} \left. \frac{\partial \Phi}{\partial x}\right |_i +\frac{\delta_{x}^2}{2!}\theta_r^2 \left. \frac{\partial^2 \Phi}
{\partial x^2}\right |_i + \frac{\delta_{x}^3}{3!} \theta_r^3 \left. \frac{\partial^3 \Phi}{\partial x^3}\right |_i+O\left(\frac{\partial^3 \Phi}{\partial x^3}\right),
\end{equation}
\begin{equation}\label{backward}
\Phi_{i-1}=\Phi_i - \theta_l \delta_{x} \left.\frac{\partial \Phi}{\partial x}\right |_i +\frac{\delta_{x}^2}{2!}\theta_l^2  \left.\frac{\partial^2 \Phi}{\partial x^2}\right |_i - \frac{\delta_{x}^3}{3!} \theta_l^3 \left.\frac{\partial^3 \Phi}{\partial x^3}\right |_i+ O\left(\frac{\partial^3 \Phi}{\partial x^3}\right).
\end{equation}
According to Eqs. (\ref{forward}) and (\ref{backward}), we can obtain
\begin{equation}\label{non2}
\begin{gathered}
  \left. \frac{\partial^2 \Phi}{\partial x^2}\right |_i= \frac{2}{\theta_l \theta_r (\theta_r + \theta_l) \delta_{x}^2} \left(\theta_l \Phi_{i+1} + \theta_{r} \Phi_{i-1} - (\theta_l+ \theta_r) \Phi_i \right) -\frac{\delta_{x}}{3}(\theta_r - \theta_l) \left. \frac{\partial^3 \Phi}{\partial x^3}\right |_i \\
   -\frac{\delta_x^2}{12}(\theta_r^2+\theta_l^2-\theta_r\theta_l) \left.\frac{\partial^4 \Phi}{\partial x^4}\right |_i
  + O\left( (\theta_r^2+\theta_l^2)(\theta_r-\theta_l)\delta_x^3 \right),
\end{gathered}
\end{equation}
and
\begin{equation}\label{non1}
 \left. \frac{\partial \Phi}{\partial x}\right |_i=\frac{\theta_l^2 \Phi_{i+1}-\theta_r^2 \Phi_{i-1}-(\theta_l^2-\theta_r^2)\Phi_i}{\theta_r\theta_l(\theta_r+\theta_l) \delta_{x}}
 -\frac{\delta_{x}^2}{6}\theta_r\theta_l \left. \frac{\partial^3 \Phi}{\partial x^3}\right |_i
 +O\left( \theta_r\theta_l \delta_x^3\right).
\end{equation}
By defining $a=\theta_l \theta_r$, $b=\theta_l+\theta_r$, $c=\theta_r- \theta_l$.
The first and second derivative values can be approximated by the following expressions,
\begin{equation}\label{nonphi2}
  \left. \frac{\partial^2 \Phi}{\partial x^2}\right |_i=
  \frac{2}{ab\delta_{x}^2} (\theta_l \Phi_{i+1} -b\Phi_i+ \theta_{r} \Phi_{i-1} )
  -\frac{\delta_x}{3}c \left.\frac{\partial^3 \Phi}{\partial x^3}\right |_i
  + O\left( (b^2-3a)\delta_x^2 \right),
\end{equation}
\begin{equation}\label{nonphi}
  \left. \frac{\partial \Phi}{\partial x}\right |_i=
  \frac{1}{ab\delta_{x}} (\theta_l^2 \Phi_{i+1} +bc\Phi_i- \theta^{2}_{r} \Phi_{i-1} )
  +O(a\delta_x^2).
\end{equation}
If $\theta_l=\theta_r$, namely uniform grid, the above discrete formulas are equivalent to the center difference with second-order accuracy. For a non-uniform grid,
if the adjacent grids are not changed drastically, i.g, $c \leq \delta_x$,
the scheme above still has at least second order accuracy on the non-uniform grid.
Analogously, the first- and second-order derivatives in two dimensions are also easily derived.
\end{appendix}


\end{document}